\documentclass[sigplan,twocolumn]{acmart}
\renewcommand\footnotetextcopyrightpermission[1]{}

\usepackage{tikz}
\usepackage{amsmath}
\usepackage[normalem]{ulem}
\usepackage{endnotes,microtype,xspace,fancyvrb,multirow}
\usepackage{xspace}
\usepackage{balance}
\usepackage{pifont}
\usepackage{anyfontsize}
\usepackage{booktabs}
\usepackage{graphicx}
\usepackage{fp}
\usepackage{caption}
\usepackage[newfloat=true]{minted}
	\setminted{
		fontsize=\small,
		baselinestretch=1.0,
		breaklines=true,
		tabsize=4,
		linenos=true,
		numbersep=8pt,
		frame=lines,
		framesep=2mm,
		autogobble=true,
		breakafter=d/,
		breakbefore=.,
        xleftmargin=10pt,
	}
\usepackage[most]{tcolorbox}
\definecolor{limitationback}{HTML}{F5F7FA}
\definecolor{limitationborder}{HTML}{AAB7C8}
\tcbset{
  mybox/.style={
    colback=limitationback,
    colframe=limitationborder,
    boxrule=0.4pt,
    top=1mm, bottom=1mm,
    before skip=5pt,
    after skip=5pt,
  }
}

\usepackage{listings}
\usepackage{balance}
\usepackage{textcomp}
\usepackage{stmaryrd}
\usepackage{enumitem}
\usepackage{array}
\usepackage{wrapfig}
\usepackage{makecell}
\usepackage{subcaption}
\usepackage{tabularx}
\usepackage{fontenc}
\usepackage{bbding}
\usepackage{url}
\usepackage{colortbl}

\usepackage{layout}
\usepackage{multicol}
\usepackage{longtable}

\usepackage{xcolor}
\usepackage{hyperref}
\hypersetup{
    colorlinks=true,
    linkcolor=blue!50!black,     %
    citecolor=violet!70!black,   %
    urlcolor=cyan!60!black       %
}
\usepackage{cleveref}

\RequirePackage[T1]{fontenc}
\RequirePackage[tt=false, type1=true]{libertine}
\RequirePackage[varqu]{zi4}
\RequirePackage[libertine]{newtxmath}

\usepackage{xparse}
\RequirePackage{comment}

\DeclareMathAlphabet{\mathpzc}{OT1}{pzc}{m}{it}
\RequirePackage[mathcal]{euscript}
\RequirePackage{bbold}
\newcommand{\sys}{\textsf{Memorix}\xspace}

\newminted{c}{gobble=2,linenos,fontsize=\small}

\definecolor{zjcolor}{RGB}{81, 158, 62}

\newcommand{\myparagraph}[1]{\vspace {2pt}\noindent\emph{\textbf{#1}}}

\definecolor{diaozi}{RGB}{93,49,49}
\definecolor{hightcode}{rgb}{1.0,0.13,0.32}
\definecolor{myRubineRed}{RGB}{209, 0, 86}

\begin{document}

\date{}

\title{\sys{}: Managing Memory Selections for AI Agents}

\author{Jing Zhang}
\affiliation{
    \institution{Shanghai Jiao Tong University}
    \city{Shanghai}
    \country{China}
}

\author{Xiaguannan Song}
\affiliation{
    \institution{Shanghai Jiao Tong University}
    \city{Shanghai}
    \country{China}
}

\author{Dong Du}
\authornote{Corresponding author: \href{mailto:dd_nirvana@sjtu.edu.cn}{\nolinkurl{dd_nirvana@sjtu.edu.cn}}.}
\affiliation{
    \institution{Shanghai Jiao Tong University}
    \city{Shanghai}
    \country{China}
}

\author{Erhu Feng}
\affiliation{
    \institution{Shanghai Jiao Tong University}
    \city{Shanghai}
    \country{China}
}

\author{Yubin Xia}
\affiliation{
    \institution{Shanghai Jiao Tong University}
    \city{Shanghai}
    \country{China}
}

\settopmatter{printacmref=false}
\settopmatter{printfolios=true}

\begin{abstract}

AI agents now handle a growing share of tasks across sessions within a persistent project.
Their behavior depends not only on the model,
but heavily on the persistent state they read from the environment.
Existing systems assemble this memory at use time,
splicing selected files into the prompt as a one-off payload.
As a result,
per-task assembly is costly and can still miss critical material,
failures cannot be traced or replayed,
and agents can read documentation written for the wrong version.
Our key insight is that what an agent should read is mostly determined by its project,
so the selection can be computed once, ahead of time, and reused across tasks.

We propose \sys,
an agent memory system that realizes this insight.
\sys achieves this via three key mechanisms:
(1) an immutable and versioned \emph{Store} pins every source by identity and reproduces historical layouts cheaply,
(2) a \emph{View} that splits selection (\emph{Derivations}) from composition (\emph{Environment}), making the result reproducible and shareable,
and (3) \emph{Quests} and \emph{Profiles} keep a delivered view refreshable and available inside the project as sources evolve.
We implement \sys and evaluate it on SWE-ContextBench and LibEvolutionEval
against mainstream agent memory systems.
Precomputed Environments raise resolved rate by up to $1.5\times$
and deliver the correct library version on every version-conditioned task,
while cutting amortized cost by up to $100\times$ and build time by up to $38\times$.

\end{abstract}

\makeatletter
\let\memorix@fnsymbol\@fnsymbol
\renewcommand{\@fnsymbol}[1]{%
  \ifnum#1=1\relax\protect\Envelope\else\memorix@fnsymbol{#1}\fi}
\maketitle
\let\@fnsymbol\memorix@fnsymbol
\makeatother
\pagestyle{plain}

\section{Introduction}
\label{s:intro}

\begin{figure}[t]
	\centering
	\includegraphics[width=\linewidth]{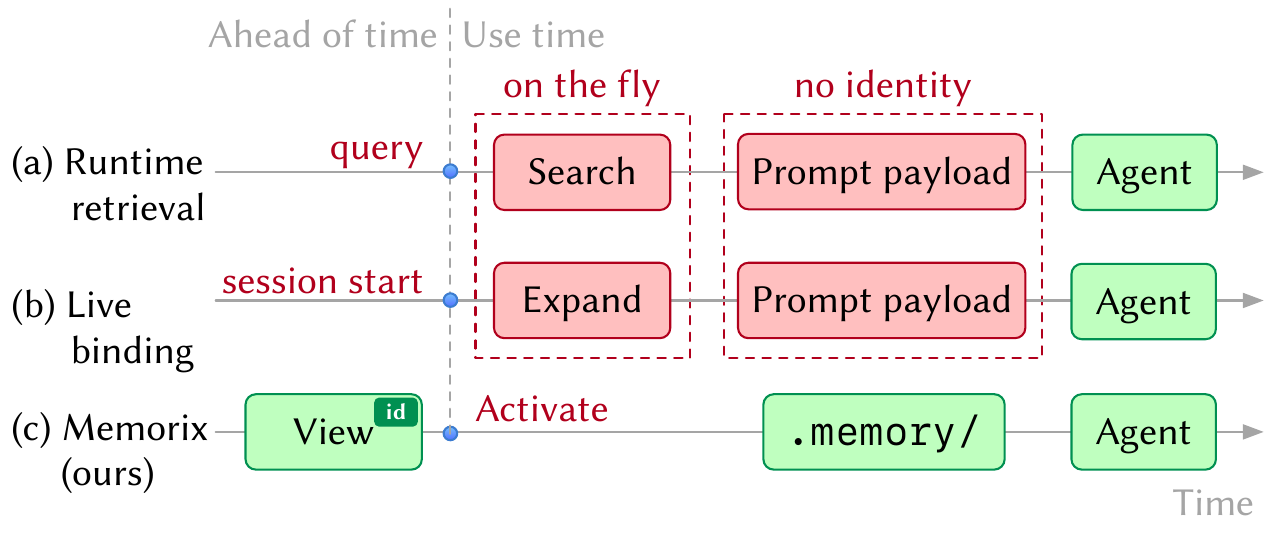}
    \caption{\textbf{Context selection at use time vs. ahead of time.}
	\emph{Runtime retrieval and live binding produce an anonymous prompt payload on the fly.
	\sys precomputes the selection as an object with identity and binds it into the project.}}
	\label{fig:intro}
\end{figure}

AI agents now perform a growing range of tasks once handled by humans~\cite{robbes2026agenticmuchadoptioncoding,watanabe2026useagenticcodingempirical,Li_2026}.
These agents reason about the task, run commands, and read the outputs before deciding the next move~\cite{yao2022react}.
Coding agents~\cite{codex, claudecode, pi}, for example,
usually run this loop over source files and test feedback.
As such work routinely spans multiple sessions~\cite{effective_harness,vasilopoulos2026codifiedcontextinfrastructureai,kwa2026measuringaiabilitycomplete}
within one persistent project directory,
an agent's behavior depends not only on the underlying model,
but heavily on the persistent contexts it reads from the environment.
We collectively refer to this persistent state,
including rule files, records of past work, and external documentation,
as \emph{agent memory}.

This memory is growing rapidly in both scale and diversity.
In coding scenarios,
the number of repositories maintaining agent rule files has grown exponentially~\cite{ORBi-3df206e9-d0b2-4a53-99ed-66d67c2946a3},
and many now keep multiple types simultaneously
--- the vast majority provide both \texttt{AGENTS.md}~\cite{openai2026agentsmd} and \texttt{CLAUDE.md}~\cite{anthropic2026claudememory}.
In office settings,
Cowork-style products~\cite{claude_cowork, codex, workbuddy, openclaw, hermes}
bring agents into daily document processing,
where the files they rely on (e.g., PDFs, Word documents, and PPTs) accumulate day by day.
Agents also generate their own records such as snapshots, patches, and tool outputs.
These records persist and shape every subsequent turn.

Agent work differs from chat in how memory is used.
In chat,
each question is new,
so what to read is decided per query.
\emph{In agent work,
one project persists across many sessions,
and what an agent should read is mostly determined by the project itself:}
the code in the repository, the versions it pins, and the prior work it continues.
Agent memory therefore has three requirements.
First,
tasks in one project largely need the same material,
so what to read should be chosen once and reused across tasks.
Second,
developers evaluate an agent's result and debug its failures by inspecting the agent's input,
so the memory that the agent read should be recorded and replayable.
Third,
the documents given to the agent should match the versions the project uses.
For example,
documents written for other versions can describe APIs that potentially conflict with the current project.

However, current agent memory systems fall short in all three respects.
First,
each task assembles what to read from scratch,
so agents spend extra retrieval steps and can still miss something entirely.
Second,
the system copies the files into the prompt and keeps no record of the input.
The agent harness may also process the same rule files in its own way,
and the files at the same paths can change.
Hence, a failure cannot be traced precisely and the same context cannot be replayed.
Third,
a memory system often holds docs for many versions,
and the system can give the agent the wrong release,
or mix several together.
Then agents can write calls against an API that does not exist in the project's version.

All three failures share one practice:
what the agent reads is assembled at use time, and the system keeps nothing of it.
It stores the sources but never the selection (i.e., which set of files and at which versions),
so the input that most shapes an agent's behavior is also the one it manages least.

Existing approaches only work around this mismatch.
\emph{Versioned files}~\cite{chacon2014gitobjects, readthedocsVersions, 10.1109/MSR.2019.00030, low2023git, li2025memos}
can persist and version each piece of content, but they cannot compose them into a single referenceable result.
\emph{Runtime retrieval}~\cite{10.1145/3448016.3457550, mem0-open-source-overview, DBLP:conf/ecai/ChhikaraKASY25, li2025memos, bytedance2026openviking, 10.14778/3827998.3828037, 3495724.3496517, karpukhin-etal-2020-dense, 10.1145/3586183.3606763, packer2023memgpt, ICLR2024_8a2acd17}
can fetch relevant content for semantic queries, but cannot guarantee consistent or deterministic results across runs.
\emph{Live binding}~\cite{openai2026agentsmd, anthropic2026claudememory, skill, openai2026buildskills, anthropic2026skills, google2026geminimd, cursor2026rules}
can declare partial relationships between source files and consumers, but the binding stops at the declaration
--- the context is actually resolved at runtime and remains anonymous.
\emph{Run snapshots}~\cite{langchain2026langgraphpersistence, ICLR2025_a4b6ad6b, dong2024agentopsenablingobservabilityllm, 270438, 203227, Ansel_2009, dong2026deltaboxscalingstatefulai}
can precisely record and replay the execution state, but the recording is bound to one run and never becomes a reusable memory object.

Our key insight is that,
\emph{since what an agent should read is mostly determined by its project,
the selection can be computed ahead of time and kept as an object}.
Like a file,
the object has an identity.
Hence, later tasks can ask for the same one by id instead of assembling the context again.
It also records what it is made of,
so the system can reproduce it,
and notice when its sources change.
Finally, the system can deliver it into the project as ordinary files.

However, realizing this abstraction introduces three technical challenges.
First, \emph{source files change continuously.}
Precomputed selections must be able to precisely reproduce historical states while also rapidly reflecting new states after source updates, without requiring a full recomputation every time.
Second, \emph{selections typically involve multi-layered sources and compositional relationships.}
The system must compose them into one identifiable result, while each constituent selection remains independently addressable.
Third, \emph{a delivered view will keep changing after the agent starts reading it.}
The system must be able to rebuild the view when sources change, and keep it available inside the project as a newer or older result takes its place.

Based on our insight, we propose \sys,
an agent memory system that computes context selection ahead of time
and represents the result as a first-class object with stable identity.
To realize this, \sys introduces three key designs to overcome those challenges.

First, \sys treats each file as an immutable object within a \emph{Store} and organizes them into a versioned logical file tree.
An edit writes a new object and leaves the old version in place,
so an earlier selection can still retrieve the object it used.
Parent-child nodes record the hierarchy among files.
When an object changes,
\sys rebuilds only the chain to the root and reuses unchanged siblings,
so a historical layout can be reproduced without rewriting the rest of the tree.
A \texttt{latest/} index then points at the current tip, so a new resolve can start from the live structure rather than walk history.
The Store therefore holds every source an agent may later read, at a pinned version that others can share by identity.

Second, \sys introduces the abstraction of \emph{View} on top of the Store.
A \emph{Derivation} selects Store objects at fixed versions and arranges them as one identifiable result.
For example, one Derivation may hold general Python guidance, and another the narrower rules for a vLLM-style repository.
An \emph{Environment} mounts one or more Derivations into a single logical tree that the agent reads as ordinary files, while each Derivation stays independently addressable.
Splitting selection from composition is what makes sharing by reference possible: two projects can mount the same Python Derivation and differ only in the project-specific mounts.
Changing what is mounted, or swapping a mounted Derivation, allocates a new Environment and leaves the previous one intact, so a run can record the exact tree it received.

Third, \sys keeps a delivered view refreshable and available
through \emph{Quests} and \emph{Profiles}.
A Quest records the intent behind a view and the identifiers the intent has resolved to,
so a later session can find that view by semantic search,
rather than selecting from the Store again.
Once any referenced Store object gains a new version,
the Quest is marked stale;
the next session to access it is notified,
and accepting the update re-resolves the intent to new Store objects.
A Profile in the project points to an Environment and, when activated, places that Environment under \texttt{.memory/}.
Moving the pointer upgrades or rolls back the project memory.
Together, Quests and Profiles turn a delivered view into a durable part of the project, even as its sources move.

We implement \sys{} as a set of CLI tools, packaged with an agent skill~\cite{skill} that teaches agents how to invoke them.
We evaluate it on SWE-ContextBench~\cite{zhu2026swecontextbenchbenchmark} and LibEvolutionEval~\cite{kuhar-etal-2025-libevolutioneval}.
Our results show that precomputing project memory as an identifiable Environment improves agent task success while remaining cheap to build and share.
For example, \sys pins library documentation to the target version on every LibEvolutionEval task, instead of mixing or missing versions at retrieval time.
On project-level coding tasks, it raises resolved rate to $64.6\%$ vs. $44.4\%$ for the no-memory agent and $59.6\%$ for the best retrieve-at-use baseline.
Amortized cost drops by up to $100\times$ relative to heavy ingest.
Build time and storage drop by up to $38\times$ and $22\times$, respectively,
because a small project-level build is reused across tasks.
Finally, we will fully open-source \sys{}.

\section{The Case for Agent Memory}
\label{s:motiv}

\subsection{Agent Execution and Memory}

In modern software development, an increasing number of tasks are handled by AI agents.
Rather than merely replying to a user’s question,
these agents invoke tools to interact with the external environment,
thereby producing higher-quality results.
In this section we explain the key concepts of agents and the persistent state they rely on.

\myparagraph{Agent execution loop.}
Large language models (LLMs) operate over a context window that is first prefilled with prior content.
The model then autoregressively generates tokens until it emits a special end-of-turn token.
To overcome the limitations of a static model (e.g., the lack of real-time information or the inability to act on the external world),
modern systems introduce tool calling.
In this setting the model may emit a sequence of special tokens that encode a function call.
Rather than returning the call to the user,
the agent runtime intercepts it,
executes the corresponding tool (e.g., a Python function, a file-system operation, or a shell command),
and appends the tool’s result to the context window.
The model then continues generation conditioned on the newly appended result.
This loop of generation, tool execution, and context extension constitutes a single agent turn
and can be repeated across multiple turns and sessions.

\myparagraph{Forms of persistent state.}
Beyond the transient contents of a single context window,
agents depend on a variety of persistent state that survives across turns and sessions.
We collectively refer to this state as \emph{agent memory}.
It typically takes three forms.
First,
rule and instruction files (e.g., \texttt{AGENTS.md}, \texttt{CLAUDE.md}, \texttt{.cursorrules})
reside inside a repository and encode style constraints, workflow conventions, and safety boundaries that the agent is expected to follow.
Second,
agents themselves leave behind artifacts during execution --- snapshots of files they have read,
generated patches, intermediate results, and tool outputs --- which are frequently re-read in subsequent sessions.
Third,
agents consult external documentation and dependency information,
such as API references and version-locked library specifications; the particular version of a document an agent reads often directly determines the version of an API it later emits.
All of these artifacts are stored as ordinary files in the project directory or in shared locations,
and they form the durable substrate on which an agent’s cross-session behavior is built.

\myparagraph{Memory resolution and delivery.}
Memory reaches the context window through several routes
that differ in \emph{when} the content is materialized and \emph{who} triggers it.
Rule files (e.g., \texttt{AGENTS.md})
are automatically resolved by the agent runtime before the loop starts
and are injected in full.
These files can further pull in other files through \texttt{@import},
which the runtime expands at load time.
Not all memory is injected upfront.
A rule file in a subdirectory takes effect only when the agent accesses files under that directory.
Skills~\cite{skill} inject only their names and descriptions as an index,
and the agent fetches the full body through a tool call when it judges them relevant.
Ordinary files in the workspace enter the context only when the agent reads them with file tools during the loop.
Other systems instead retrieve memory fragments from external stores by similarity search at run time.

\subsection{From Chat to Agent}

Traditional chat systems operate under high query dynamism.
Each user turn introduces a new semantic intent,
so the relevant context is typically determined at runtime
through similarity search or on-demand expansion.
\emph{Agent workloads differ in a fundamental way.}
An agent usually executes multi-turn, multi-session tasks inside one persistent workspace.
What it needs to read is largely shaped by the repository it belongs to,
the pinned versions of its dependencies,
and the artifacts left by its own earlier work.
These relationships are mostly determined by the project structure
rather than by the semantics of any single task.
As a result, the sources that enter an agent’s context exhibit far greater stability across turns and sessions than those in conventional chat settings.

\subsection{Limitations of Agent Memory}

Three structural limitations hinder current agent memory.

\begin{figure}[t]
	\centering
	\includegraphics[width=0.99\linewidth]{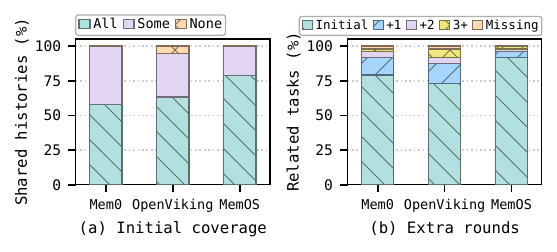}
	\setlength{\belowcaptionskip}{-10pt}
	\setlength{\abovecaptionskip}{-10pt}
	\caption{\textbf{Shared memory lacks a stable selection.}
    \emph{Across 19 histories linked to 48 tasks:
   (a)~per history, first-recall coverage over its linked tasks;
   (b)~per task, extra memory-access rounds before its linked history arrives,
    including \emph{Missing} tasks that never receive it.}}
	\label{fig:m1-selection}
\end{figure}

\begin{table}[t]
    \centering
    \footnotesize
	\setlength{\belowcaptionskip}{3pt}
	\setlength{\abovecaptionskip}{-12pt}
    \setlength{\tabcolsep}{3pt}
    \renewcommand{\arraystretch}{1.15}
    \resizebox{\columnwidth}{!}{%
    \begin{tabular}{@{}lcccccccc@{}}
    \toprule
    \rowcolor{gray!18}\textbf{Probe} & \multicolumn{2}{c}{codex / swap model} & \multicolumn{2}{c}{claude / swap doc} & \multicolumn{3}{c}{deepseek / swap harness} & \multicolumn{1}{c}{no doc} \\
     & \makecell{gpt-\\5.6-sol} & \makecell{deepseek-\\v4-flash} & \makecell{CLAUDE} & \makecell{AGENTS} & \makecell{opencode} & \makecell{pi} & \makecell{aider\\read} & \makecell{aider\\auto} \\
    \midrule
    \texttt{indent} & \textcolor{teal}{\textbf{4/4}} & \textcolor{teal}{\textbf{4/4}} & \textcolor{teal}{\textbf{4/4}} & \textcolor{orange!85!black}{1/4} & \textcolor{teal}{\textbf{4/4}} & \textcolor{teal}{\textbf{4/4}} & \textcolor{teal}{\textbf{4/4}} & \textcolor{red!75!black}{0/4} \\
    \texttt{no-os.system} & \textcolor{teal}{\textbf{4/4}} & \textcolor{teal}{\textbf{4/4}} & \textcolor{teal}{\textbf{4/4}} & \textcolor{teal}{\textbf{4/4}} & \textcolor{teal}{\textbf{4/4}} & \textcolor{teal}{\textbf{4/4}} & \textcolor{teal}{\textbf{4/4}} & \textcolor{teal}{\textbf{4/4}$^{\dagger}$} \\
    \texttt{js-header} & \textcolor{teal}{\textbf{4/4}} & \textcolor{teal}{\textbf{4/4}} & \textcolor{teal}{\textbf{4/4}} & \textcolor{red!75!black}{0/4} & \textcolor{teal}{\textbf{4/4}} & \textcolor{teal}{\textbf{4/4}} & \textcolor{teal}{\textbf{4/4}} & \textcolor{red!75!black}{0/4} \\
    \texttt{chkpoint} & \textcolor{teal}{\textbf{4/4}} & \textcolor{teal}{\textbf{4/4}} & \textcolor{teal}{\textbf{4/4}} & \textcolor{red!75!black}{0/4} & \textcolor{teal}{\textbf{4/4}} & \textcolor{teal}{\textbf{4/4}} & \textcolor{orange!85!black}{1/4} & \textcolor{red!75!black}{0/4} \\
    \texttt{glob-version} & \textcolor{teal}{\textbf{4/4}} & \textcolor{teal}{\textbf{4/4}} & \textcolor{teal}{\textbf{4/4}} & \textcolor{orange!85!black}{2/4} & \textcolor{teal}{\textbf{4/4}} & \textcolor{teal}{\textbf{4/4}} & \textcolor{teal}{\textbf{4/4}} & \textcolor{red!75!black}{0/4} \\
    \texttt{final-marker} & \textcolor{teal}{\textbf{4/4}} & \textcolor{teal}{\textbf{4/4}} & \textcolor{teal}{\textbf{4/4}} & \textcolor{red!75!black}{0/4} & \textcolor{teal}{\textbf{4/4}} & \textcolor{teal}{\textbf{4/4}} & \textcolor{orange!85!black}{3/4} & \textcolor{red!75!black}{0/4} \\
    \texttt{test-first} & \textcolor{teal}{\textbf{4/4}} & \textcolor{teal}{\textbf{4/4}} & \textcolor{teal}{\textbf{4/4}} & \textcolor{orange!85!black}{3/4} & \textcolor{orange!85!black}{2/4} & \textcolor{orange!85!black}{2/4} & \textcolor{red!75!black}{0/4} & \textcolor{red!75!black}{0/4} \\
    \texttt{yaml-priority} & \textcolor{red!75!black}{0/4} & \textcolor{red!75!black}{0/4} & \textcolor{red!75!black}{0/4} & \textcolor{red!75!black}{0/4} & \textcolor{red!75!black}{0/4} & \textcolor{red!75!black}{0/4} & \textcolor{red!75!black}{0/4} & \textcolor{red!75!black}{0/4} \\
    \texttt{indent-conflict} & \textcolor{teal}{\textbf{4/4}} & \textcolor{teal}{\textbf{4/4}} & \textcolor{teal}{\textbf{4/4}} & \textcolor{teal}{\textbf{4/4}} & \textcolor{teal}{\textbf{4/4}} & \textcolor{teal}{\textbf{4/4}} & -- & \textcolor{teal}{\textbf{4/4}$^{\dagger}$} \\
    \bottomrule
    \end{tabular}%
    }
    \caption{\textbf{Same rules, different obedience.}
    \emph{Rows are nine rule-following tasks; columns are eight ways of delivering the same rules. Each cell counts how many of four runs obeyed. ``--'' marks cells where the base task itself failed. $\dagger$ marks cases where the model already complied without seeing the rules. Rules and model stay fixed across columns.}}
    \label{tab:m2-behavior}
\end{table}

\myparagraph{Unstable selection.}
Current agent memory systems retain shared memory,
but each use retrieves a fresh subset of it.
Retrieval ranks fragments against the current task's query,
so the same history surfaces differently for each task.
Related tasks in one project therefore have no stable selection to reuse.
\autoref{fig:m1-selection}
shows the consequence on SWE-ContextBench~\cite{zhu2026swecontextbenchbenchmark}.
Even the strongest (MemOS~\cite{li2025memos}) puts only 79\% of shared memory into every related task's first recall (Mem0~\cite{mem0-open-source-overview} 58\%, OpenViking~\cite{bytedance2026openviking} 63\%),
so part of the shared memory does not reach some related tasks.
The natural fix is to ask again,
but raising coverage needs extra memory-access rounds
(one model response that initiates memory access),
and 8\%--27\% of related tasks need at least one.
\autoref{fig:m1-selection}(b) shows that some tasks still receive nothing at all.
As a result,
related tasks keep paying repeated fetch cost,
and shared memory can still be missed entirely (details in \autoref{s:eval}).

\begin{tcolorbox}[mybox]
\textbf{Limitation 1:}
Without a stable selection, each task re-fetches the shared memory, and some of it still never arrives.
\end{tcolorbox}

\myparagraph{Unrecorded delivery.}
Current agent memory systems hand the agent an assembled context,
but keep no record of the input.
\autoref{tab:m2-behavior} shows the costs of this in practice.
We use nine tasks,
each requiring one specific rule to complete,
and deliver the same rule file in eight ways,
with one model swap as a control.
It shows three key findings:
\emph{(1) Default injection decides success.}
Claude Code ignores \texttt{AGENTS.md},
so renaming the file causes the majority of js-header, chkpoint, and final-marker runs to fail.
\emph{(2) The editing protocol can conflict with a correct rule.}
Aider~\cite{aider-edit-formats} expects a filename plus a fenced code block,
but in indent-conflict the model emitted valid JSON that obeyed the documented rule,
so the task still failed.
\emph{(3) Natural-language priorities are unreliable.}
No harness respected the highest-priority rule in yaml-priority.
Retrieve-at-use systems have the same gap:
each use builds a fresh context,
leaving nothing stable to replay.
File-path delivery fails similarly.
Since the content under a path can change,
it does not identify what the agent actually read.

\begin{tcolorbox}[mybox]
\textbf{Limitation 2:}
Without a record of the input, a failure cannot be traced and a delivery cannot be replayed.
\end{tcolorbox}

\begin{figure}[t]
	\centering
	\includegraphics[width=0.95\linewidth]{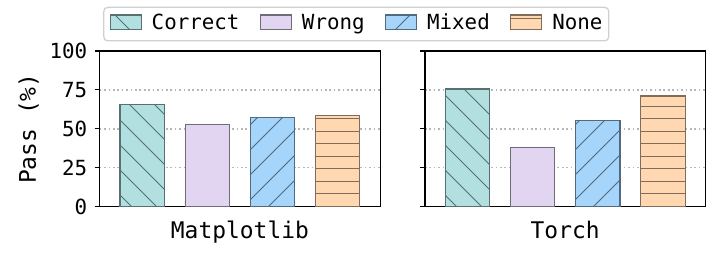}
	\setlength{\belowcaptionskip}{-10pt}
	\setlength{\abovecaptionskip}{0pt}
	\caption{\textbf{Wrong-version documents hurt accuracy.}
    \emph{Pass rate on LibEvolutionEval~\cite{kuhar-etal-2025-libevolutioneval} when the agent receives only the required-version documents (Correct),
    only other versions (Wrong), both (Mixed), or no documents (None).}}
	\label{fig:m3-documents}
\end{figure}

\myparagraph{Unversioned delivery.}
A project usually uses one version of a library,
but the memory pool holds documents for many.
At each use,
the system recalls some of them without pinning the choice to the project version.
\autoref{fig:m3-documents} shows why that pin matters.
With only correct documents,
agents pass 65.9\% of Matplotlib tasks and 75.6\% of Torch tasks.
With only wrong documents, pass rates fall to 52.7\% and 37.8\%.
Mixed documents also lag behind correct-only documents,
and wrong documents can hurt more than giving no documents at all
(Matplotlib 52.7\% vs. 58.7\%, Torch 37.8\% vs. 71.1\%).
Mismatched APIs will enter the context unnoticed,
and the agent can write calls against an API that does not exist in the project's version
(details in \autoref{ss:version}).

\begin{tcolorbox}[mybox]
\textbf{Limitation 3:}
Without pinned versions, recall cannot guarantee documents that match the project's version.
\end{tcolorbox}

\subsection{Current Works}

\newcommand{\cmark}{\textcolor{green!60!black}{\ding{51}}}	%
\newcommand{\xmark}{\textcolor{red}{\ding{55}}}				%
\newcommand{\wmark}{\textcolor{orange}{\ding{115}}}			%
\newcommand{\approachref}[2]{%
  #1\par
  \vspace{1.5pt}%
  {#2}%
}
\begin{table*}[t]
\centering
\setlength{\belowcaptionskip}{5pt}
\setlength{\abovecaptionskip}{-5pt}
\resizebox{\textwidth}{!}{%
\begin{tabular}{@{}lcccccc@{}}
\toprule
\parbox[b][2\baselineskip][b]{1.8cm}{\raggedright\textbf{Approach}}
& \parbox[b]{2.6cm}{\centering Version individual\\sources}
& \parbox[b]{3.8cm}{\centering Declarative multi-source\\composition}
& \parbox[b][2\baselineskip][b]{2.1cm}{\centering Selection\\identity}
& \parbox[b]{2.0cm}{\centering Delivery\\attribution}
& \parbox[b]{1.8cm}{\centering Freshness\\tracking}
& \parbox[b]{1.9cm}{\centering Workspace\\binding}
\\
\midrule
\approachref{Versioned files}{~\cite{chacon2014gitobjects, readthedocsVersions, 10.1109/MSR.2019.00030, low2023git, li2025memos}}
& \cmark & \xmark & \xmark & \xmark & \wmark & \wmark \\
\addlinespace
\approachref{Runtime retrieval}{~\cite{10.1145/3448016.3457550, mem0-open-source-overview, DBLP:conf/ecai/ChhikaraKASY25, li2025memos, bytedance2026openviking, 10.14778/3827998.3828037, 3495724.3496517, karpukhin-etal-2020-dense, 10.1145/3586183.3606763, packer2023memgpt, ICLR2024_8a2acd17}}
& \wmark & \xmark & \xmark & \xmark & \wmark & \xmark \\
\addlinespace
\approachref{Live binding}{~\cite{openai2026agentsmd, anthropic2026claudememory, skill, openai2026buildskills, anthropic2026skills, google2026geminimd, cursor2026rules}}
& \cmark & \wmark & \xmark & \xmark & \xmark & \wmark \\
\addlinespace
\approachref{Run snapshots}{~\cite{langchain2026langgraphpersistence, ICLR2025_a4b6ad6b, dong2024agentopsenablingobservabilityllm, 270438, 203227, Ansel_2009, dong2026deltaboxscalingstatefulai}}
& \wmark & \xmark & \wmark & \cmark & \xmark & \xmark \\
\addlinespace
\midrule
\textbf{\sys (Ours)}
& \cmark & \cmark & \cmark & \cmark & \cmark & \cmark \\
\bottomrule
\end{tabular}%
}
\caption{\textbf{Existing mechanisms for agent memory.}
    \emph{Each row is a class of current practice.
Columns are capabilities that agent memory needs.
\cmark{} means the capability is supported.
\xmark{} means it is not.
\wmark{} means support is partial or only at an adjacent layer.
    Existing classes cover individual slices, but none of them satisfy all.}}
\label{tab:comparison}
\end{table*}

\myparagraph{Versioned files.}
Git~\cite{chacon2014gitobjects, low2023git} and other systems~\cite{10.1109/MSR.2019.00030,li2025memos}
keep each document on its own history,
so a checkout can restore a past state of those files.
Rule files (e.g., \texttt{AGENTS.md}~\cite{openai2026agentsmd}, \texttt{CLAUDE.md}~\cite{anthropic2026claudememory}) live in the same layout,
and documentation hosts likewise publish a separate build for each release~\cite{readthedocsVersions}.
Reproducing one file, or even a full checkout, is therefore straightforward.
However,
agent memory needs something more:
the files a project wants the agent to read together form a chosen subset,
and file-level history alone does not turn that subset into a unit the project can point to,
reuse, or watch when one of its inputs changes.

\myparagraph{Runtime retrieval.}
Systems (e.g., Mem0~\cite{mem0-open-source-overview, DBLP:conf/ecai/ChhikaraKASY25}, OpenViking~\cite{bytedance2026openviking, 10.14778/3827998.3828037}, and MemOS~\cite{li2025memos})
retrieve fragments at use time by similarity or graph lookup~\cite{10.1145/3448016.3457550, 3495724.3496517, karpukhin-etal-2020-dense, ICLR2024_8a2acd17}.
That design fits chat,
where each turn can change what matters
and the useful text may sit anywhere in a large corpus~\cite{10.1145/3586183.3606763, packer2023memgpt}.
But agent work is different.
The agent stays inside one project,
so what it should read follows from that project
rather than from the wording of a single query.
Because each use builds a fresh subset on the spot,
two related tasks can receive different fragments
even when the rules and the model stay the same.
Related tasks therefore do not share one durable subset they can rely on,
and a change in an upstream input leaves nothing clear to revisit.

\myparagraph{Live binding.}
Other systems declare which sources an agent may read
and expand those links when the session starts
or when the agent later reaches the matching path.
Imports~\cite{anthropic2026claudememory, google2026geminimd}, directory-scoped rules~\cite{openai2026agentsmd, anthropic2026claudememory, google2026geminimd, cursor2026rules}, and skill indexes~\cite{skill, openai2026buildskills, anthropic2026skills} all work this way.
They keep the underlying files in ordinary project history
and they state part of the relationship up front.
The declaration, however, stops before delivery:
what the runtime actually puts into the prompt is still assembled on the spot
and left without a durable handle.
The same files can therefore yield a different prompt under a different harness,
with nothing left to replay that delivery
or to tell whether an input has since changed.

\myparagraph{Run snapshots.}
A run snapshot freezes one execution so that run can resume~\cite{Ansel_2009} or play back.
For example, LangGraph checkpointers keep the messages and tool outputs at a given step~\cite{langchain2026langgraphpersistence}.
Similar record-and-replay~\cite{270438, 203227} or sandbox tools~\cite{dong2026deltaboxscalingstatefulai} do the same for a whole environment.
Although this record answers what one agent saw~\cite{ICLR2025_a4b6ad6b,dong2024agentopsenablingobservabilityllm},
the next agent still needs to know what to read in this project.
A frozen thread does not answer that,
and it will not reflect later edits to those sources either.

\myparagraph{Summary.}
As shown in \autoref{tab:comparison},
each existing mechanism already covers one slice of the problem.
But
\emph{none of them treats the selected context as an object with its own identity.}
Related tasks therefore rebuild that context for each use,
and the project has little left to replay or pin afterward.
Agent work makes a better option available,
because what should be read usually follows from the project itself.
\sys{} computes the selection ahead of time
and therefore covers the gap.

\section{\sys Overview}
\label{s:overview}

\begin{figure}[t]
	\centering
    \setlength{\belowcaptionskip}{-15pt}
    \setlength{\abovecaptionskip}{5pt}
	\includegraphics[width=0.8\linewidth]{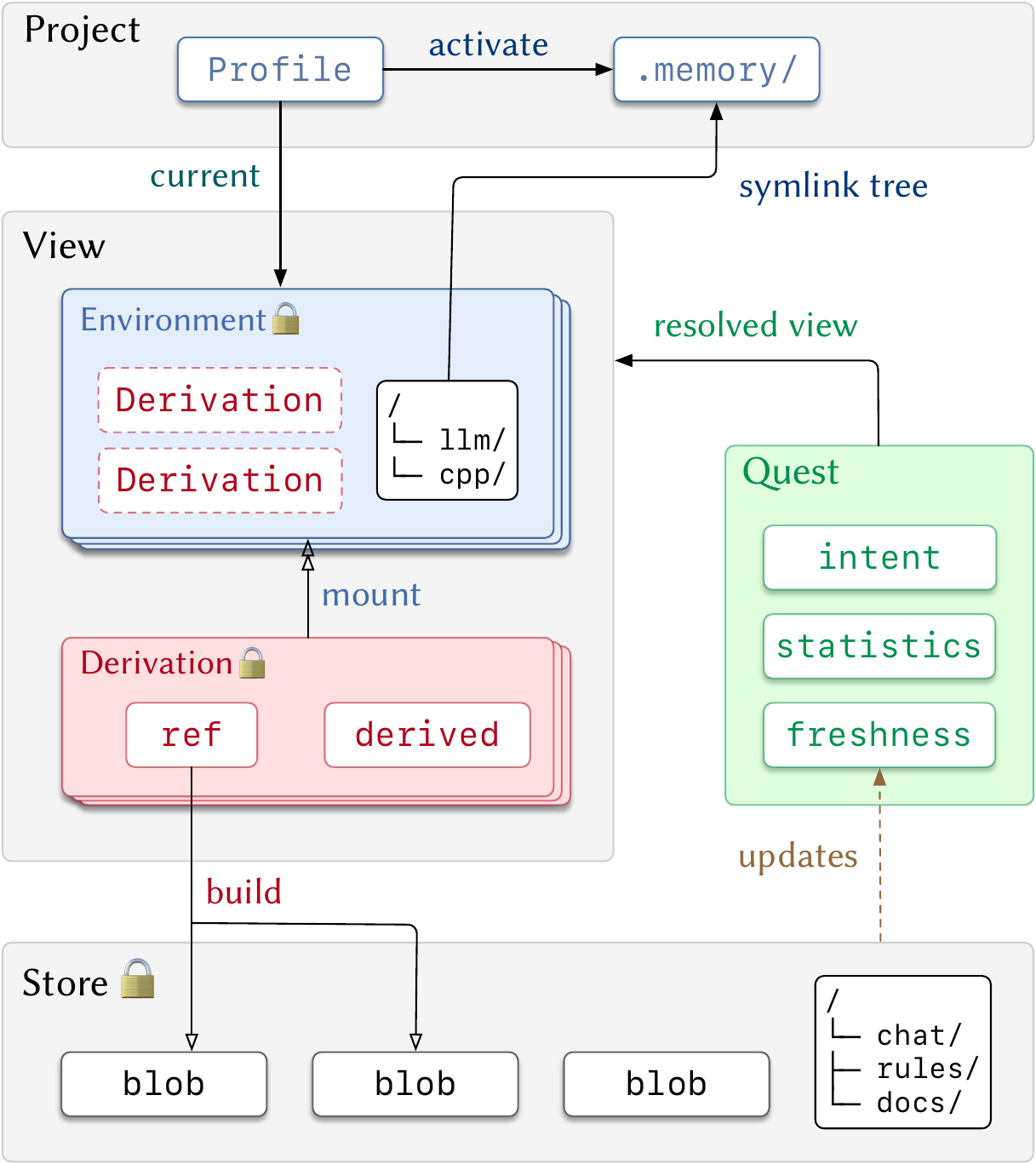}
    \caption{\textbf{Overview of \sys{}.}
    \emph{Locked objects are immutable: a Derivation pins Store blobs and may carry derived artifacts, and an Environment mounts Derivations as one tree.
    Each resolved view is recorded on a Quest that tracks intent, statistics, and freshness. Store updates mark the Quest stale.
    A Profile points at the current Environment and activates it as a symlink tree under \texttt{.memory/}.}}
	\label{fig:overview}
\end{figure}

\subsection{Insight}

The core limitation is that selected context never becomes an object.
Related work therefore either stores raw sources and resolves them at use time, or searches for fragments when a query arrives.
Both follow from the same assumption: what the agent should read will change with the next question.
Our observation is that agent work drops this assumption.
The agent stays inside one project,
so the sources it should read are mostly determined by that project.
Thus,
our insight is that
\emph{context selection can be precomputed and explicitly represented as an object with identity.}
Later work can point at that identity instead of assembling the same context again.

\subsection{Object Model and Architecture}

\sys{} makes every computed result an immutable object with a stable identity,
following the object model of Nix~\cite{269552}.
The context an agent reads is one such result.
\sys{} introduces five object types:
\emph{Store, Derivation, Environment, Quest, and Profile.}
We call a Derivation or an Environment a \emph{view}: a precomputed selection with a stable identity.
\autoref{fig:overview} shows how they refer to each other.

\myparagraph{Store.}
\sys{} starts with files that later steps can refer to by identity.
Each remembered file is an immutable object.
An edit creates a new object and leaves the old one in place,
so an earlier identity is still addressable.
Objects sit in one tree whose root is global memory.
The top level of that tree is memory kind.
For instance,
\texttt{/rules} for written instructions,
\texttt{/docs} for project or API documents,
and \texttt{/chat} for traces taken from earlier runs.
Users then add their own levels under each kind.
For example, \texttt{/rules/python} can hold the general Python guidance,
while \texttt{/rules/python/vllm} contains guidance for vLLM-style repositories only.
The tree records which paths exist and which versions each path has.
A \texttt{latest/} index points at the current tip of every path,
so a new resolve can start from the tips rather than walk history.

\myparagraph{Derivation.}
Once files have identities, a selection of them must have an identity as well.
A Derivation is that object.
It lists Store objects at pinned versions and arranges them in a subtree,
so resolving the same Derivation yields the same tree.
Different Derivations can select different Store objects.
One may hold Python-project memory, another C++ rules, another the docs for library $L$ at version $v$.
A Derivation may also carry derived artifacts of that selection: a summary of the chosen files, or embeddings used for retrieval over those files.
Those artifacts hang on the Derivation itself.

\myparagraph{Environment.}
A project often needs several Derivations at once,
but they should appear as one tree
so that the agent can browse and read that context with ordinary file tools.
An Environment mounts each Derivation into that tree and is the object later delivered to an agent.
For example,
one Environment may place
a Python Derivation at \texttt{/python}
and a C++ Derivation at \texttt{/cpp}.
Changing the mount set, or changing a mounted Derivation, allocates a new Environment, and the previous one remains addressable.

\myparagraph{Quest.}
Source files keep changing after a view exists,
so it must be able to report when its inputs have moved and to be resolved again.
A Quest records why a Derivation or an Environment was built: the task the view serves, and the files it set out to gather.
\sys{} writes the Quest in the same step as the view,
so the intent stays attached to the objects it produced.
In addition,
the Quest is text, and an agent can grep it.
Each resolve returns one Derivation identifier or one Environment identifier,
and the Quest keeps a history of those identifiers.
When a Store object under the current resolve gains a new version,
the Quest marks that resolve stale and reports to the user.
A later resolve can then produce a new identifier,
once the user accepts the update.
The Quest also counts resolves and reads as demand on the view.
That signal can prompt a new summary on the Derivation or a split into finer Derivations.

\myparagraph{Profile.}
The Environment is what the agent should read,
but it is still an object outside the project.
An agent works in a project directory and needs that Environment to appear there,
under a binding that can move.
A Profile is that binding:
a mutable pointer, kept in the project, to one Environment identifier.
Moving the pointer to a newer Environment upgrades the memory.
Moving it to an older identifier rolls the memory back.
After the pointer is set, \sys{} builds a tree of symbolic links
under the project directory (e.g., \texttt{.memory/}).
Hence,
the agent can read that directory with ordinary file tools,
and a run records the Environment identifier as the context it received.

\subsection{Workflow}

The end-to-end flow has two paths: ingest and resolve.

\myparagraph{Ingest.}
A file enters the Store as an immutable object under its path.
If that path already holds a file, the edit creates a new version and leaves the previous one addressable.
Also, it updates the freshness of Quests that hold the old version.

\myparagraph{Resolve.}
To obtain a view, the agent first greps Quest text for a matching intent.
A fresh match is reused directly.
If the match is marked stale, the user decides whether to refresh it.
If no suitable view exists,
the agent walks the Store tree,
selects the needed files,
and builds Derivations.
The selected Derivations are then mounted into an Environment.
Building each Derivation or Environment also creates a Quest
that records the intent behind that view.
A Profile in the project then points at the Environment.
When the Profile is activated,
\sys{} can materialize the Environment under the project (e.g., \texttt{.memory/}),
and the agent reads Store objects through that tree using ordinary file tools.

\section{Detailed Design}
\label{s:design}

\subsection{Store: Versioned Memory Tree}
\label{ss:store}

\begin{figure}[t]
	\centering
	\includegraphics[width=\linewidth]{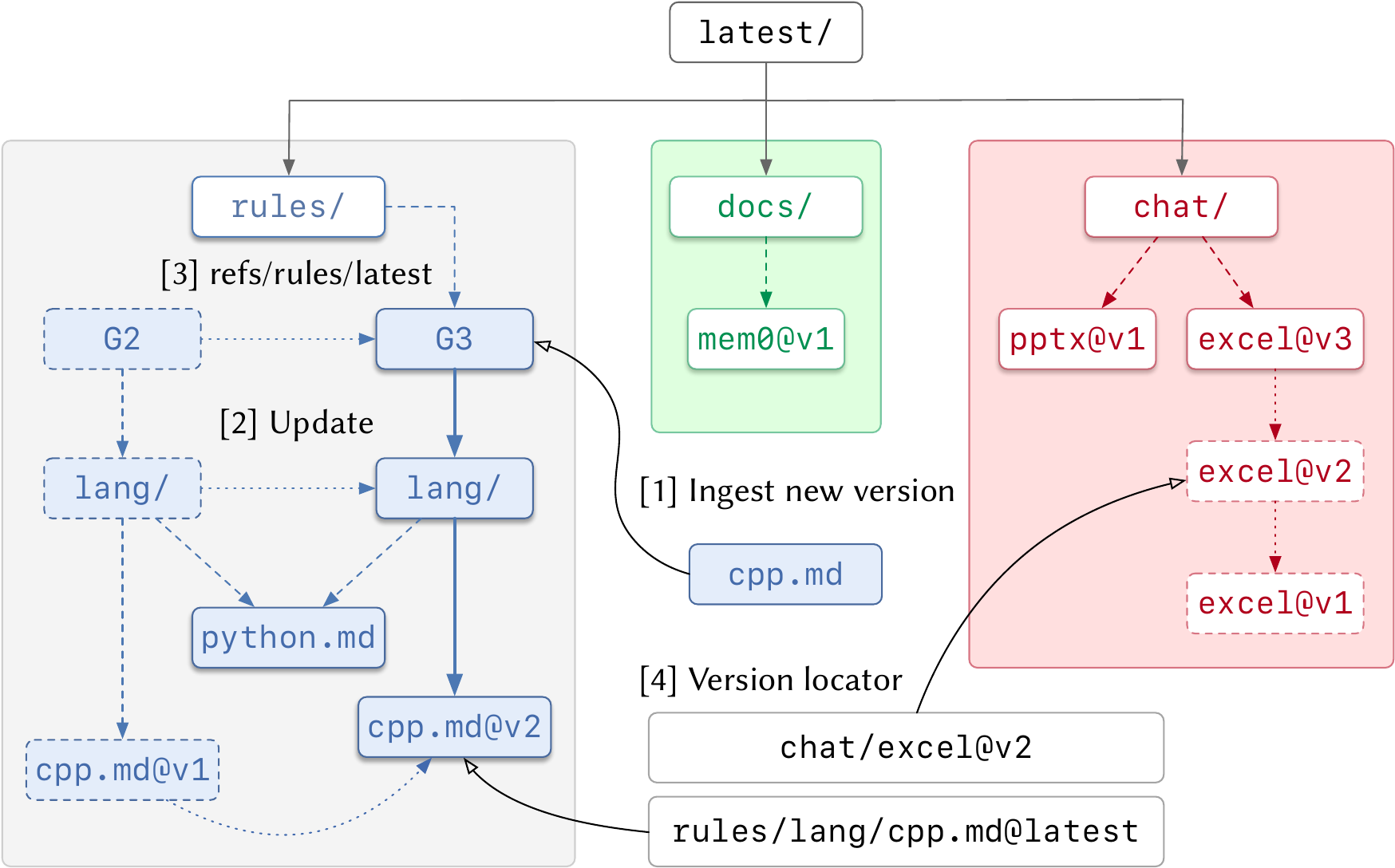}
	\setlength{\belowcaptionskip}{-5pt}
	\setlength{\abovecaptionskip}{-5pt}
	\caption{\textbf{The Store.}
	\texttt{latest/} is the checkout of typed trees.
	An ingest of a new \texttt{cpp.md} writes a blob, \texttt{cpp.md@v2} ([1]).
	The update rebuilds only the ancestor chain: \texttt{lang/} is rewritten and generation G3 succeeds G2, while \texttt{python.md} is reused and the other trees do not move ([2]).
	\texttt{refs/rules/latest} then advances to G3 ([3]).
	A version locator such as \texttt{rules/lang/cpp.md@latest} points to the new version, while older ones (e.g., \texttt{chat/excel@v2}) remain addressable ([4]).}
	\label{fig:store}
\end{figure}

Because source files keep changing,
\sys{} must record each modification and retrieve a version efficiently.
As \autoref{fig:store} shows,
\sys addresses this with a versioned \emph{Store}.

\myparagraph{Immutable objects.}
The Store must be able to give back the same file content that was previously used.
In practice,
projects usually keep one file at each path and overwrite it whenever it changes.
After an overwrite, however, the path no longer refers to the old content,
so the earlier selection cannot be reproduced.
A path can locate the current file,
but we need a way to locate the content itself.
Hence, when \sys admits a file on \texttt{ingest},
it computes the SHA-256 of the content
and stores the content at \texttt{blobs/<hash>/object}.
The blob is immutable,
so a later change becomes a new \texttt{ingest} under a new hash,
while the old hash still retrieves the old content.
Similarly,
a deletion marks the file without removing the content.
A selection records blob hashes and reproduces itself by reading the blobs at those hashes.

\myparagraph{Bottom-up versioning.}
Content is stored by hash, yet each blob is still placed at a path.
The Store must keep a versioned layout of those path-to-blob placements, and update it cheaply when only one path changes.
A common idea is to save each new layout as a full listing of all paths and blobs, rewriting unchanged paths along with the rest.
As paths accumulate, each small edit therefore becomes slow.
We observe that \emph{an edit usually touches only the chain from the changed file up to the root, while neighboring directories keep the same children.}
Therefore,
\sys records each directory as an immutable tree object
and rebuilds only that chain bottom-up (a Merkle tree~\cite{646752.704751}).
A tree object is a \texttt{trees/<hash>.json} that maps each child name to a blob hash or to another tree hash.
When a leaf gets a new blob,
\sys writes new trees for its ancestors and reuses old tree hashes for unchanged siblings.
An \texttt{ingest} records a new generation that points at the new root tree,
and links to the previous one.
Reading a generation walks that root tree to the blob hashes,
so an earlier layout can still be reproduced.

\myparagraph{Typed trees.}
The Store holds written rule files, project documents, and traces from earlier runs.
These files are written by different people and change on different schedules.
If we keep them in one tree and tell them apart by folder names,
every edit advances a single generation chain,
which is inefficient and costly.
Since these types do not depend on each other,
\sys gives each type its own tree:
\texttt{rules/}, \texttt{docs/}, and \texttt{chat/}.
Each tree holds its own blobs, tree objects, and generations.
An \texttt{ingest} writes the blob into the target tree
and records a new generation of that tree.
The other trees do not change.
An \texttt{rm} records a new generation of that tree with the path gone.

\myparagraph{Latest tip.}
Once layouts are recorded as generations,
the Store must still make the current layout efficient to reach.
Generations link only backward, so nothing in the graph points at the current state;
ordinary tools such as \texttt{ls} and \texttt{cat} expect files on a path,
not a generation graph.
Once the current generation is known,
however,
its tree already lists every path and blob,
so it suffices to remember that generation as the tip
and present those files on ordinary paths.
Hence,
\sys keeps a movable reference at \texttt{refs/<type>/latest}
that records the current generation
and maintains a read-only hardlink tree under \texttt{latest/}
that mirrors that generation's paths onto blob objects.
An \texttt{ingest} that updates a path advances the generation,
moves the reference, and refreshes those hardlinks.
Historical layouts remain reachable by generation id.

\myparagraph{Locators.}
A current file sits on a path under \texttt{latest/},
while an older file is reached by generation id.
A selection has to pick the tree, the path, and either the tip or a generation.
\sys records that as a \emph{locator}.
For example,
\texttt{rules/cpp.md} selects one file
and \texttt{docs/api/} selects a subtree,
and either can take \texttt{@latest} or \texttt{@<generation-id>}.
\texttt{@latest} resolves to the file currently shown under \texttt{latest/}.
An ingest destination is the same form without a version selector.

\subsection{Views: Derivation and Environment}

\begin{figure}[t]
	\centering
	\includegraphics[width=\linewidth]{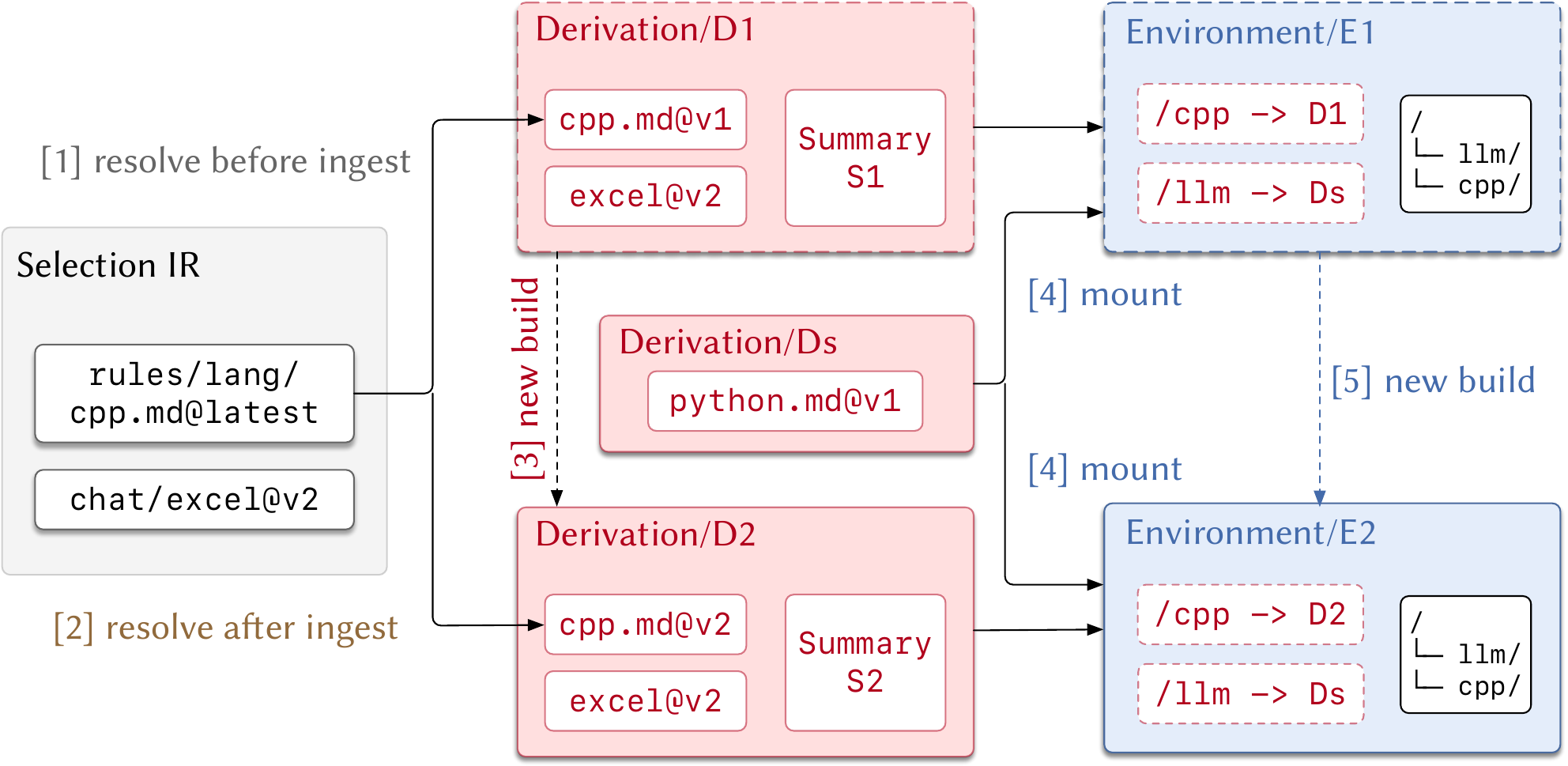}
	\Description{One selection IR resolves to Derivation D1 before an ingest and to D2 after it, while excel stays pinned at v2. Each Derivation carries a summary. E1 and E2 mount D1 and D2 at /cpp and share Ds at /llm. Dashed borders mark earlier results.}
	\setlength{\belowcaptionskip}{-5pt}
	\setlength{\abovecaptionskip}{-5pt}
	\caption{\textbf{Building and composing Views.}
	The same selection IR resolves to D1 before an ingest ([1]) and to D2 after it ([2]).
	The locator \texttt{rules/lang/cpp.md@latest} advances from v1 to v2, while \texttt{chat/excel@v2} stays pinned.
	The new build ([3]) produces D2 with a fresh summary, and mounting ([4]) places D1 or D2 at \texttt{/cpp} and the shared Ds at \texttt{/llm}, so swapping D1 for D2 creates E2 ([5]).}
	\label{fig:views}
\end{figure}

Selections involve multi-layered sources and compositional relationships, so they must form one identifiable result while each part stays independently addressable.
\sys addresses this with \emph{Views}: \emph{Derivations} and \emph{Environments}.
\autoref{fig:views} shows how a Store update changes one selection while a shared View remains reusable.

\myparagraph{Selection IR.}
As discussed in \autoref{ss:store},
a locator can take \texttt{@latest} or a generation id;
a selection is a list of such locators,
so some entries move when the Store moves.
Resolving a selection yields files,
but the result no longer records which locators produced it,
so it cannot be re-evaluated after the Store changes.
\sys therefore keeps the unresolved list as a \emph{selection IR}.
The IR stores each locator's tree, path, and version selector.
A \texttt{build} evaluates it against the Store:
\texttt{@latest} reads the latest ref,
a generation id opens that generation,
and each path then walks to blobs.
For example,
an IR with \texttt{rules/cpp.md@latest} and \texttt{docs/api/@generation/03}
reads the current rules generation for the first locator
and generation \texttt{03} of documents for the second,
then takes \texttt{cpp.md} and every file under \texttt{api/}.
After a new ingest of \texttt{rules/cpp.md},
the same IR picks up the new rules blob while still selecting the same \texttt{api/} files.

\myparagraph{Derivation.}
A selection of Store files must remain referable as one result, including a fixed folder the agent can read.
A common idea is to keep a list of pinned \texttt{path@generation} entries, and to walk that list and build a folder whenever the files are needed.
The list can still be edited in place, so the selection as a whole can change even when every \texttt{path@generation} is pinned.
We observe that, \emph{as with file content in the Store, the selection itself needs an immutable object.}
Hence, \sys records each built selection as an immutable \emph{Derivation}.
A \texttt{build} writes the pinned generation and blob for every chosen path into \texttt{resolved\_inputs},
hashes them into a \texttt{derivation\_id}, and stores the ready folder under \texttt{tree/}.
The same resolved inputs yield the same id,
so later steps can mount or open that Derivation and read the same folder.
A Derivation may also carry derived artifacts of that selection (e.g., a summary).
\sys{} stores them keyed by the derivation id,
so every project that mounts the same Derivation reuses them without recomputing.

\myparagraph{Environment.}
A project needs one memory folder the agent can read,
while parts of that memory must stay reusable in other projects that overlap only in part.
If we put everything a project needs into one large Derivation,
shared files and project-only files then become one selection.
Then a second project cannot take the shared part without the project-only part,
and derived artifacts on the shared files cannot be reused on their own.
Therefore,
the shared part and the project-only part must be separate selections.
\sys records a composition of Derivations as an immutable \emph{Environment}.
An Environment stores \texttt{bindings} that mount each Derivation at a path,
and hashes them into an \texttt{environment\_id}.
It presents one folder by placing each Derivation's \texttt{tree/} under its mount path.
Changing what is mounted, or swapping a mounted Derivation, creates a new Environment.
Finally,
each Derivation id remains independently usable,
so shared selections and their derived artifacts can be mounted into many Environments.

\myparagraph{Conflict resolution.}
When an Environment mounts several selections, their contents may disagree, so \sys needs an explicit rule for which file wins.
There are many kinds of conflicts.
For example,
a file may have several versions,
two rules may sit at different depths of the same path family,
or guidance from different memory kinds may conflict.
Therefore, \sys resolves each case as follows.
For multiple versions of the same file,
it prefers the latest version,
or the version locked by the repository when a lock is present.
For conflicts inside one path family
(e.g., general Python rules vs. vLLM rules),
it prefers the deeper, more specific leaf.
For conflicts across memory kinds,
it orders by trust:
\texttt{/rules} over \texttt{/docs} over \texttt{/chat},
because hand-written rules are the most precise,
documents describe the codebase less authoritatively,
and chat extracts are the noisiest.
Finally,
these rules are only defaults,
and users can override these defaults explicitly.

\subsection{Quest and Profile: Evolution and Delivery}

\begin{figure}[t]
	\centering
	\includegraphics[width=\linewidth]{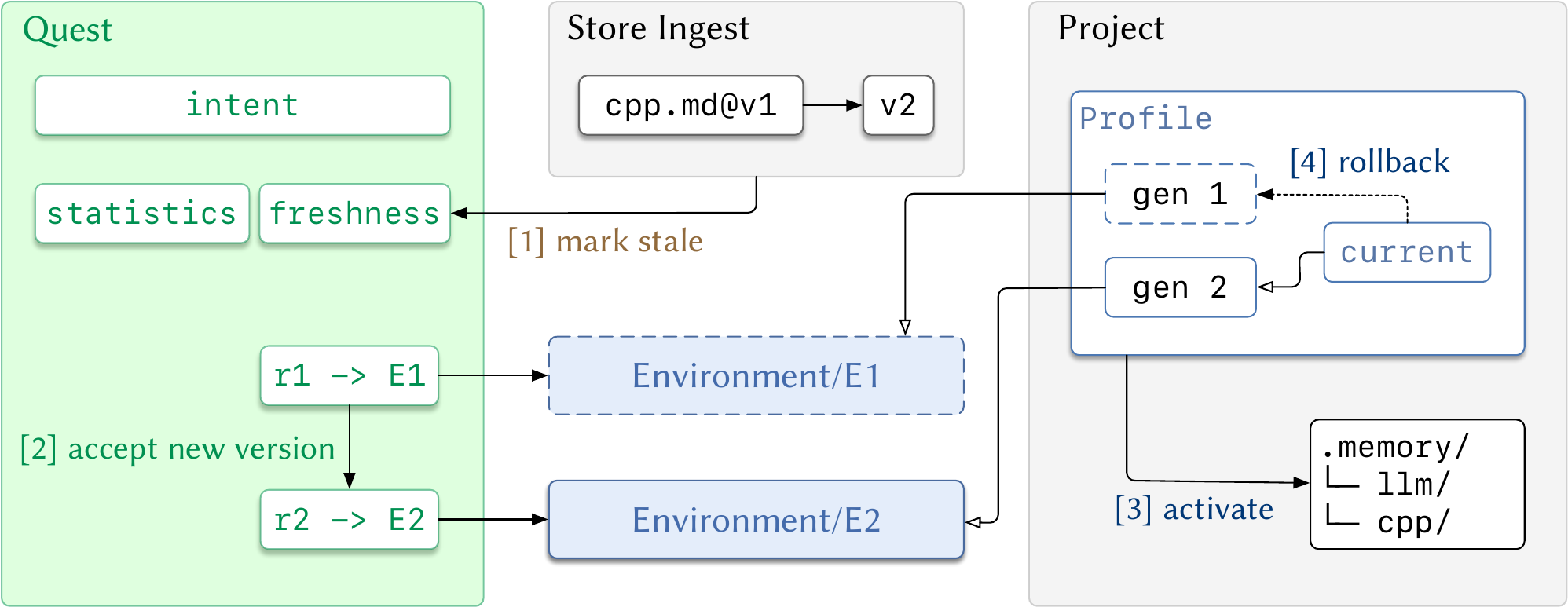}
	\Description{A Store ingest marks the Quest's freshness stale. Accepting the new version appends resolve r2 pointing at E2 while r1 and E1 stay addressable. Activation sets the Profile's current generation to gen 2, which points at E2, and refreshes the symlink tree at .memory/. Moving current back to gen 1 rolls back to E1. Quest results and Profile generations refer to the same immutable Environments.}
	\setlength{\belowcaptionskip}{-5pt}
	\setlength{\abovecaptionskip}{-5pt}
	\caption{\textbf{Evolving and delivering a view.}
	A Store ingest marks the dependent Quest stale ([1]).
	Accepting the new version appends resolve r2 to the Quest history ([2]).
	Activation sets the Profile's current generation to gen 2 and refreshes the symlink tree ([3]).
	Moving \texttt{current} back to gen 1 rolls the project memory back to E1 ([4]).}
	\label{fig:lifecycle}
\end{figure}

A delivered view must be findable again later,
stay fresh as the Store moves,
and remain available inside the project as a newer or older result takes its place.
\sys addresses this with \emph{Quests} and \emph{Profiles}.
\autoref{fig:lifecycle} separates the history of resolved views from the versions activated in a project.

\myparagraph{Quest intent.}
After a view exists,
a later session or project must be able to find a suitable one again
without selecting from the Store from scratch.
A naïve way is to let an agent browse the pool of Views,
and search the selected source files to decide which view fits.
That agentic check can improve accuracy,
but verifying many candidates this way becomes prohibitive as the pool grows.
We observe that \emph{whether a view fits depends on the task it serves},
which a view's selected files do not record.
Hence,
\sys writes a \emph{Quest} whenever a view is built.
The Quest stores searchable intent text
and links to the Derivation or Environment that view produced.
Agents discover Quests with \texttt{find} or \texttt{grep} over that intent,
then open the linked view.
Each later \texttt{resolve} appends the new result id to the Quest's history,
so the same intent keeps a trail of resolved views over time.

\myparagraph{Freshness.}
A Quest's current resolve can fall behind as the Store gains new versions.
A simple method is to rebuild on every use,
but it wastes work when nothing relevant has changed.
Another is to keep the old Environment with no check,
but it leaves the agent on expired context with no signal.
\sys{} maintains a reverse index from Store paths to Quests,
so whenever an ingest completes,
it marks every Quest stale
whose current resolve still depends on the old version.
When a later session finds this Quest, it sees that mark.
If the user wants to accept that update,
\sys{} re-evaluates the corresponding View,
creates a new View,
and appends it to the Quest history,
while the old history remains in the Quest so that the user can see the changes.

\myparagraph{Profile binding.}
The Environment lives in the Store,
but the agent works inside a project directory
and needs that memory as ordinary files under the project.
Copying the files into the repository
loses the clean binding to the Environment,
risks inconsistency with the Store, and is hard to refresh or roll back.
\sys therefore keeps a \emph{Profile} in the project that points to one Environment.
It \texttt{activates} an Environment under the project as a symlink tree at \texttt{.memory/}.
The agent reads with ordinary file tools,
and a run can record the Environment id as the context it received.

\myparagraph{Upgrade and rollback.}
After a Profile is bound,
the project must be able to switch to a newer or older Environment
without the user tracking ids by hand.
Re-running activate with a new Environment id moves the project forward,
but leaves rollback to out-of-band bookkeeping.
Hence,
\sys keeps an ordered binding history in the Profile,
with a \texttt{current} pointer to one Environment.
Moving current forward selects a newer binding and upgrades the memory;
moving it back selects an older one and rolls the memory back.
Each move refreshes the symlink tree under \texttt{.memory/}
to match the Environment now selected.

\section{Implementation}
\label{s:imp}

We have implemented a \sys prototype in 30k lines of C++,
together with a skill document that agents use to drive the CLI.
This section covers the implementation choices that matter most for usability and correct delivery.

\myparagraph{Usability.}
Users prefer to build and search memory in conversation:
for example, asking an agent to put a document into memory.
Precise control of \sys, however, goes through a CLI that takes exact identifiers
(e.g., path, generation)
and runs them as given.
Between these two sits a gap:
natural language is too loose for exact memory operations,
while the CLI alone is too rigid for ordinary use.
\sys closes this gap in two ways.
First, it ships the CLI manual as a skill.
The harness keeps only the skill description in the system prompt.
The agent reads the full manual on demand,
so ordinary turns stay free of the manual
while natural-language requests can still trigger exact CLI calls.
Second, \sys records each view's intent in a Quest.
The Quest gives the agent searchable semantic information
with which to find and select a suitable memory when building a new view.
The skill supplies \emph{how} to invoke the CLI.
The Quest supplies \emph{which} view fits the current intent.
Once an Environment is activated,
the agent can read it with ordinary file tools.

\myparagraph{Symlink and hardlink.}
\sys picks a different link kind for each job.
For the Store, it fills \texttt{latest/} with hardlinks.
A hardlink shares the blob's inode, so \texttt{latest/} exposes that content directly, with no extra copy.
Paths under \texttt{latest/} also look like ordinary files to tools such as \texttt{ls} and \texttt{cat}.
Refreshing \texttt{latest/} sits off the agent's read path, so the cost of \texttt{link} and \texttt{unlink} on ingest is acceptable.
The Profile needs a different binding.
Upgrade and rollback must retarget the project tree cheaply.
Besides, unlike the Store where blobs and \texttt{latest/} always share one filesystem,
a Profile may bind a project to an Environment on another filesystem.
Hence,
we use symlinks for these scenarios.
Symlinks have limits, though:
some systems disable them,
and a symlink can redirect reads outside the Store.
We leave a portable delivery mechanism to future work.

\myparagraph{Atomic updates.}
Agents read \texttt{latest/} and \texttt{.memory/} while ingest and activate rewrite those trees.
A torn tip would mix old and new paths in one turn, so \sys publishes each update atomically.
On ingest, it first writes the new blob, tree objects, and generation as immutable files.
Only then does it swing \texttt{refs/<type>/latest} to the new generation id
and replace \texttt{latest/} by building a fresh hardlink tree aside and renaming it into place.
On activate, upgrade, or rollback, it writes the Profile's \texttt{current} pointer
and replaces \texttt{.memory/} the same way: stage a new symlink tree, then rename.
Readers therefore always see either the previous tree or the new one, never a half-applied mix.
Immutable history stays reachable by generation id or by an earlier Profile generation,
so a swing of the tip does not erase what earlier resolves already pinned.

\section{Evaluation}
\label{s:eval}

\begin{figure*}[t]
	\centering
	\includegraphics[width=0.99\linewidth]{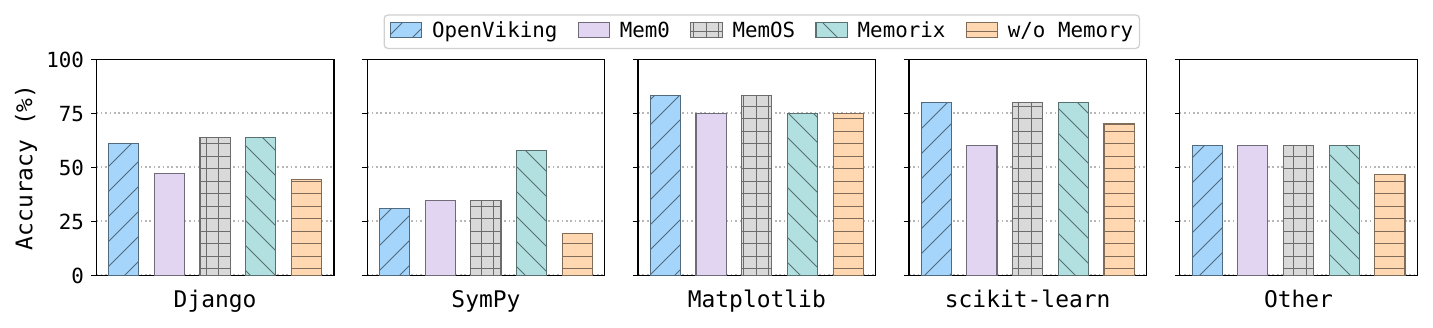}
	\setlength{\abovecaptionskip}{-3pt}
	\caption{\textbf{Resolved rate on SWE-ContextBench Lite by project group.}
	\emph{Four projects with at least 10 tasks are shown separately; \emph{Other} groups the remaining eight projects (15 tasks).}}
	\label{fig:eval-q1}
\end{figure*}

Our evaluation aims to answer the following questions:

\begin{itemize}[noitemsep,topsep=0pt,leftmargin=1em]
	\item \textit{Q1}: Does precomputed project memory improve task success relative to retrieve-at-use memory systems? (\autoref{ss:accuracy})
	\item \textit{Q2}: Does \sys lower per-task cost as more tasks share one project memory? (\autoref{ss:cost})
	\item \textit{Q3}: Does \sys reduce end-to-end latency at runtime relative to retrieve-at-use memory? (\autoref{ss:latency})
	\item \textit{Q4}: Does version-pinned memory improve correctness on version-conditioned tasks, and how do correct versus wrong version retrievals compare to baselines? (\autoref{ss:version})
	\item \textit{Q5}: What build and storage overhead does \sys introduce relative to retrieve-at-use baselines? (\autoref{ss:overhead})
\end{itemize}

\myparagraph{Benchmarks.}
SWE-ContextBench~\cite{zhu2026swecontextbenchbenchmark}
evaluates whether coding agents can reuse experience across related tasks.
We use SWE-ContextBench Lite, which builds an experience pool from agent trajectories or summaries on the 300 SWE-bench Lite~\cite{jimenez2024swebench} tasks
and adds 99 related tasks identified through relations between their GitHub issues and PRs.
LibEvolutionEval~\cite{kuhar-etal-2025-libevolutioneval}
evaluates whether models can generate API completions for a specified library version.
We select 212 version-sensitive tasks in which the required APIs differ across library versions.
Then we provide test cases for our evaluation.
For each task, using an outdated API fails the tests,
while only the API corresponding to the target version passes.
This subset therefore enables a more reliable evaluation of version-specific memory retrieval.

\myparagraph{Baselines.}
We compare \sys against three mainstream memory systems that span common organization paradigms:
OpenViking v0.4.19~\cite{bytedance2026openviking} (filesystem-based memory),
Mem0 v2.0.20~\cite{mem0-open-source-overview, DBLP:conf/ecai/ChhikaraKASY25} (vector-store memory),
and MemOS v2.0.33~\cite{li2025memos} (graph-structured memory),
plus a no-memory agent.
All systems ingest the same historical materials.
OpenViking, MemOS, and Mem0 retrieve online per task, while \sys precomputes Store objects and Derivations and binds one Environment per project.

\myparagraph{Setup.}
We use deepseek-v4.1-flash~\cite{deepseek-v41-technical-report-2026} with high thinking effort as the model and the Pi coding agent v0.85.1~\cite{pi} as the harness.
We price LLM tokens at \$0.006 / \$0.30 / \$1.20 per million tokens~\cite{deepseek-api-pricing-2026} for cached input, uncached input, and output, respectively.
Embeddings use text-embedding-3-large~\cite{openai-text-embedding-3-large} at \$0.13 per million tokens.

\subsection{Task Success}
\label{ss:accuracy}

This subsection asks whether precomputed project memory raises resolved rate on project-level coding tasks relative to retrieve-at-use memory and to running without memory.

\myparagraph{Methodology.}
We evaluate on SWE-ContextBench Lite: 99 tasks across 12 projects, with 300 past experience trajectories as the shared memory material.
OpenViking, MemOS, Mem0, and \sys each organize that material.
For \sys, we provide the material to a Pi agent equipped with the \sys skill.
The agent constructs a set of Derivations and composes them into one Environment per project.
The no-memory agent receives none.
Comparing to no-memory shows whether past experience helps at all.
Comparing to other baselines shows whether a precomputed project Environment influences accuracy.
At runtime, Pi accesses OpenViking, MemOS, and Mem0 through system-specific plugins that expose their retrieval interfaces.
For \sys, it activates the corresponding Environment under \texttt{.memory/} for Pi to access using ordinary file tools, with no extra plugin loaded at runtime.
The official SWE-bench harness then applies the patch they output and scores it.
A task is resolved when every \texttt{FAIL\_TO\_PASS} test passes and no \texttt{PASS\_TO\_PASS} test regresses.

\myparagraph{Results.}
\autoref{fig:eval-q1} shows resolved rate by project group.
Over all 99 tasks, \sys resolves $64.6\%$, ahead of MemOS ($59.6\%$), OpenViking ($57.6\%$), Mem0 ($50.5\%$), and no-memory ($44.4\%$).
Every memory system beats no-memory, so past experience indeed helps the agent generate better patches.
\sys widens that gap the most, at $+20.2$ points over no-memory and $+5.1$ points over the strongest retrieve-at-use baseline (MemOS).
The largest lift is on SymPy, where \sys reaches $57.7\%$ against $30.8$--$34.6\%$ for the retrieve-at-use systems and $19.2\%$ without memory.
On Django, \sys ties MemOS at $63.9\%$.
On Matplotlib, OpenViking and MemOS lead at $83.3\%$, while \sys matches Mem0 and no-memory at $75.0\%$.
These results show that precomputation is feasible for project-level agent work.
Binding memory into an Environment ahead of time
sustains high resolve quality under every new task.

\subsection{Cost Analysis}
\label{ss:cost}

This subsection asks whether \sys lowers per-task cost as more tasks share one project memory.

\begin{figure}[t]
	\centering
	\includegraphics[width=0.99\linewidth]{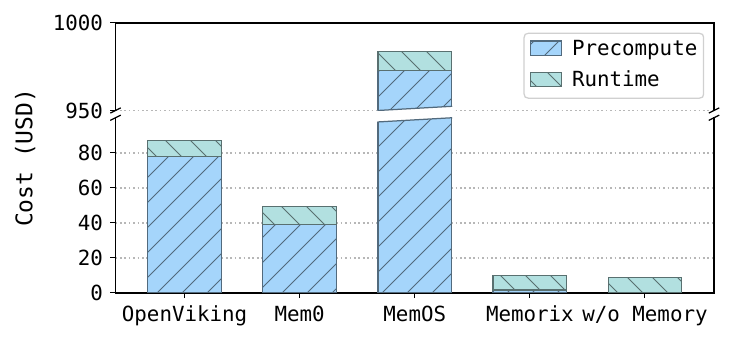}
	\setlength{\belowcaptionskip}{-10pt}
	\setlength{\abovecaptionskip}{-15pt}
	\caption{\textbf{Precompute and runtime cost.}
	\emph{Total API cost for all 99 tasks, which consists of a one-time precomputation over shared memory and runtime across all task runs.}}
	\label{fig:eval-cost}
\end{figure}

\myparagraph{Cost.}
\autoref{fig:eval-cost} splits total spend on the 99 Lite tasks into precompute and runtime.
Runtime cost is similar across systems, from $\$8.18$ for \sys to $\$10.99$ for MemOS.
Precompute dominates the gap.
MemOS spends $\$973$ before any task runs, OpenViking $\$77.7$, and Mem0 $\$39.3$.
\sys spends $\$1.66$ on precompute, so its total $\$9.83$ stays close to no-memory ($\$8.80$).
Per task, that is $\$0.099$ for \sys, about $5\times$ cheaper than Mem0 ($\$0.498$), $9\times$ cheaper than OpenViking ($\$0.877$), and $100\times$ cheaper than MemOS ($\$9.94$).

\begin{figure}[t]
	\centering
	\includegraphics[width=0.99\linewidth]{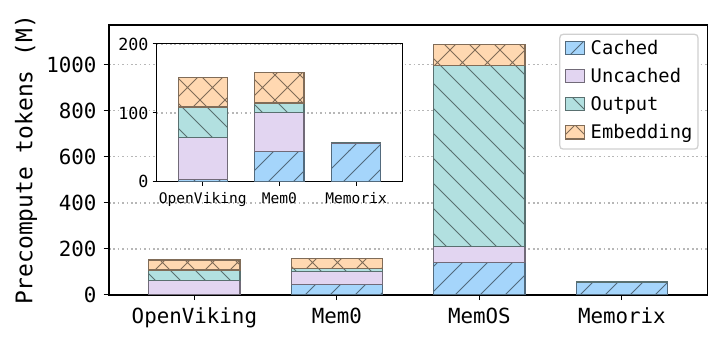}
	\setlength{\belowcaptionskip}{-10pt}
	\setlength{\abovecaptionskip}{-15pt}
	\caption{\textbf{Precompute token breakdown by kind.}}
	\label{fig:eval-token}
\end{figure}

\myparagraph{Token breakdown.}
To further analyze the cost,
\autoref{fig:eval-token} breaks precompute tokens into multiple types.
MemOS generates $783$M output tokens while building memory, about $880\times$ \sys's $0.89$M output, which accounts for most of its $\$973$ precompute bill under output pricing.
OpenViking and Mem0 spend $61$M and $57$M uncached input tokens ($70\times$ and $66\times$ \sys) plus $43$M--$44$M extra embedding tokens.
\sys uses $55.4$M cached input tokens and under $1$M each of uncached input and output.
\emph{Cached input is cheap relative to uncached input and output},
so \sys's precompute stays at $\$1.66$ even though its cached volume is non-trivial.
Runtime dollars stay close across systems because the 99 solving passes do comparable LLM work once memory is in place.

\begin{figure}[t]
	\centering
	\includegraphics[width=0.99\linewidth]{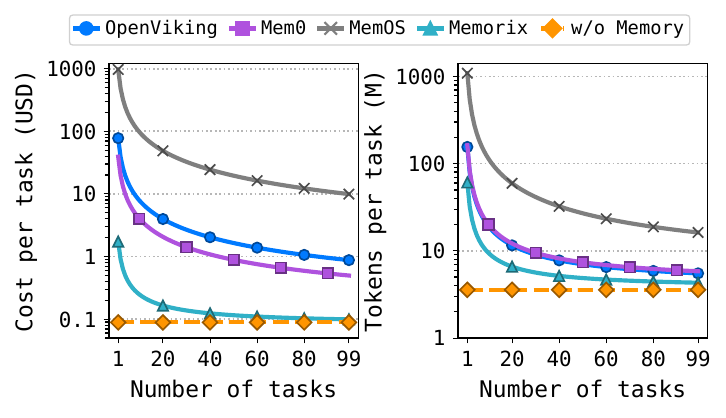}
	\setlength{\belowcaptionskip}{-10pt}
	\setlength{\abovecaptionskip}{-15pt}
	\caption{\textbf{Amortized cost and tokens per task as more tasks share one project memory.}}
	\label{fig:eval-amort}
\end{figure}

\myparagraph{Amortization.}
\autoref{fig:eval-amort} amortizes both dollars and tokens over more tasks that share one project memory.
At 10 tasks, \sys is already $\$0.25$ per task, against $\$4.03$ for Mem0, $\$7.86$ for OpenViking, and $\$97.4$ for MemOS.
At 99 tasks the figures fall to $\$0.099$, $\$0.498$, $\$0.877$, and $\$9.94$, while amortized tokens fall to $4.31$M per task for \sys versus $5.52$M--$16.2$M for the retrieve-at-use systems and $3.57$M without memory.
No-memory stays flat at $\$0.089$/task.
The curves fall because each new task reuses the same project build and mostly pays runtime.
\sys starts from a small build, so that sharing reaches near no-memory dollars and tokens, while heavier ingest baselines stay costly even at 99 tasks.

\subsection{End-to-End Latency}
\label{ss:latency}

This subsection asks whether \sys reduces end-to-end runtime latency relative to retrieve-at-use memory.

\begin{figure}[t]
	\centering
	\includegraphics[width=0.99\linewidth]{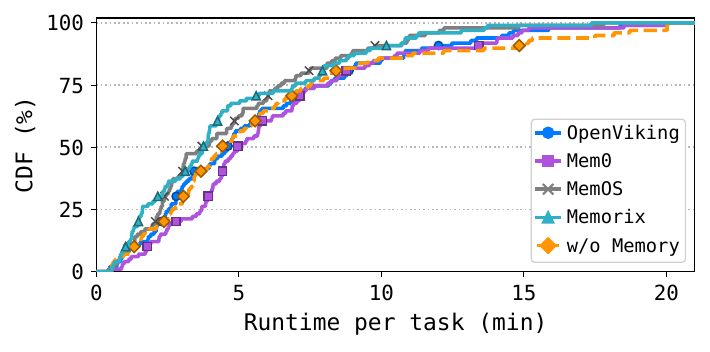}
	\setlength{\belowcaptionskip}{-10pt}
	\setlength{\abovecaptionskip}{-15pt}
	\caption{\textbf{CDF of end-to-end per-task runtime.}}
	\label{fig:eval-latency}
\end{figure}

\myparagraph{Results.}
\autoref{fig:eval-latency} shows the CDF of end-to-end latency on the same 99 Lite tasks.
Median latency is close across systems: $3.76$~min for \sys and $3.70$~min for MemOS, against $4.43$~min without memory, $4.62$~min for OpenViking, and $4.96$~min for Mem0.
The gap widens in the tail.
At p95, \sys finishes in $11.2$~min, about $1.3\times$ faster than OpenViking ($14.3$~min) and Mem0 ($14.6$~min), and $1.6\times$ faster than no-memory ($17.6$~min).
At p99, \sys is fastest at $13.8$~min, ahead of MemOS ($14.9$~min) and about $1.3\times$--$1.5\times$ faster than OpenViking, Mem0, and no-memory.
End-to-end time is still dominated by solving, which is why medians stay close.
A precomputed Environment replaces retrieval by lightweight file operations,
which cuts the long tail at runtime.

\subsection{Multi-Version Memory}
\label{ss:version}

This subsection asks whether version-pinned memory improves correctness on version-conditioned tasks, and how correct versus wrong version retrievals compare to baselines.

\myparagraph{Methodology.}
We evaluate on a 212-task subset of LibEvolutionEval:
version-specific code completion for Matplotlib ($167$) and Torch ($45$),
scored by executable tests we add.
Each task provides a code context and a target library version;
the memory system under evaluation provides the documents.
A model then fills in the missing snippet from this prompt.
A task passes when the completed code passes all tests.
\sys{} first ingests all documents into the Store (each version of a library becomes a different generation),
and then builds the Derivations and Environments.
Before each task, we use a Pi agent equipped with the \sys{} skill to select an Environment.
This step simulates a user preparing an Environment for the task's codebase.
After that, we append the contents of that Environment to the prompt.
Other systems ingest the same documents
with their native importers
and retrieve once per task using provided task information.

\begin{figure}[t]
	\centering
	\includegraphics[width=0.99\linewidth]{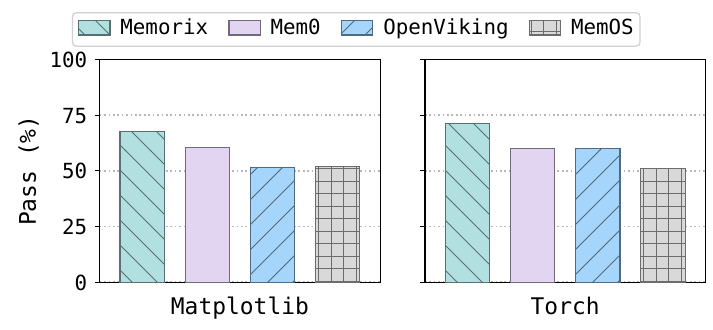}
	\setlength{\belowcaptionskip}{-12pt}
	\setlength{\abovecaptionskip}{-13pt}
	\caption{\textbf{Pass rate on the LibEvolutionEval subset.}}
	\label{fig:eval-multi}
\end{figure}

\myparagraph{Accuracy.}
\autoref{fig:eval-multi} compares pass rate on the 212 tasks.
\sys leads at $68.4\%$, about $1.1\times$ Mem0 ($60.4\%$) and $1.3\times$ OpenViking ($53.3\%$) and MemOS ($51.9\%$).
Matplotlib, the larger slice, follows that pattern at $67.7\%$.
On Torch the gap widens: \sys reaches $71.1\%$, about $1.2\times$ Mem0 and OpenViking ($60.0\%$ each) and $1.4\times$ MemOS ($51.1\%$).

\begin{figure}[t]
	\centering
	\includegraphics[width=0.99\linewidth]{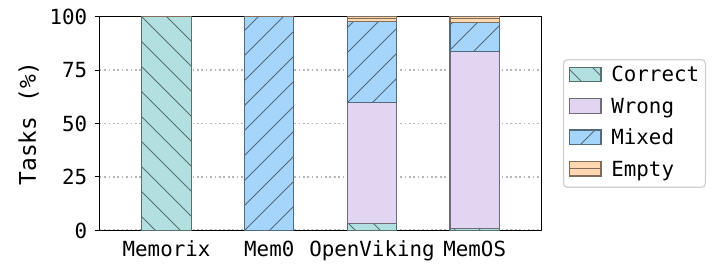}
	\setlength{\belowcaptionskip}{-10pt}
	\setlength{\abovecaptionskip}{-15pt}
	\caption{\textbf{Version composition of delivered documentation.}
	\emph{Each bar groups all 212 tasks by the source versions.}}
	\label{fig:eval-version}
\end{figure}

\myparagraph{Version reading.}
Further,
\autoref{fig:eval-version} classifies each task by the library versions that appear in the delivered documents:
\emph{correct} (only the target version), \emph{wrong} (only other versions), \emph{mixed} (target and others), or \emph{empty}.
\sys delivers only correct documents on all tasks.
Mem0 is mixed on every task.
OpenViking is correct on only $3.3\%$, wrong on $56.6\%$, and mixed on $37.7\%$.
MemOS is even worse: $0.9\%$ correct and $82.5\%$ wrong.
The accuracy gap tracks this breakdown.
Hence, a version-pinned Environment keeps the target docs in place, so the agent can follow the right API.

\subsection{Overhead Analysis}
\label{ss:overhead}

This subsection asks what build and storage overhead \sys introduces relative to retrieve-at-use baselines.

\myparagraph{Build time.}
\autoref{tab:eval-build} shows cumulative build time over the 12 projects in SWE-ContextBench.
\sys finishes in $2.2$~h.
OpenViking takes $8.4$~h (about $3.8\times$), Mem0 $27.2$~h (about $12\times$), and MemOS $84.5$~h (about $38\times$).
The long Mem0 and MemOS builds come from heavy per-item ingest work, while \sys pays a short project-level construction pass.

\begin{table}[t]
	\centering
	\footnotesize
	\setlength{\belowcaptionskip}{3pt}
	\setlength{\abovecaptionskip}{-10pt}
	\renewcommand{\arraystretch}{1.15}
	\begin{tabular}{@{}lcccc@{}}
	\toprule
	\rowcolor{gray!18}\textbf{System} & \textbf{\sys} & \textbf{OpenViking} & \textbf{Mem0} & \textbf{MemOS} \\
	\midrule
	Build time (h) & \textbf{2.2} & 8.4 & 27.2 & 84.5 \\
	\bottomrule
	\end{tabular}
	\caption{\textbf{Cumulative memory build time.}}
	\label{tab:eval-build}
\end{table}

\begin{figure}[t]
	\centering
	\includegraphics[width=0.99\linewidth, trim=0 5 0 0, clip]{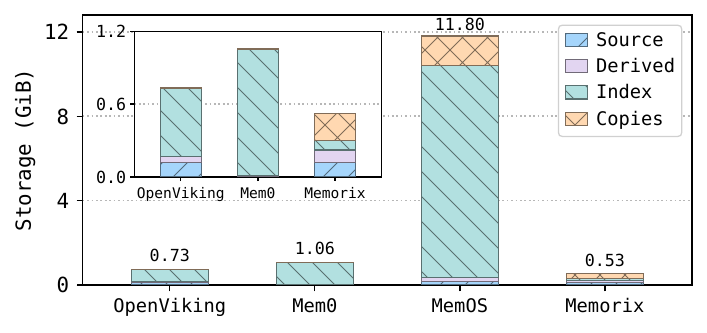}
	\setlength{\belowcaptionskip}{-5pt}
	\setlength{\abovecaptionskip}{-10pt}
	\caption{\textbf{On-disk memory footprint.}
	\emph{Disk space used by each system's persistent deployment after the run.}}
	\label{fig:eval-storage}
\end{figure}

\myparagraph{Storage.}
\autoref{fig:eval-storage} breaks down on-disk size after the build.
\sys uses $0.53$~GiB in total, about $1.4\times$ less than OpenViking ($0.73$~GiB), $2.0\times$ less than Mem0 ($1.06$~GiB), and $22\times$ less than MemOS ($11.8$~GiB).
MemOS is dominated by intermediate index and metadata (about $10.2$~GiB) plus $1.4$~GiB of serving memory.
Mem0 likewise stores most of its footprint as index metadata.
\sys keeps intermediate state small ($0.18$~GiB) and materializes $0.22$~GiB of serving memory for bound Environments.
A light project-level build therefore stays cheap in both time and space.

\section{Related Work}
\label{s:relwk}

\myparagraph{Long-horizon agents.}
Long-horizon agent work is attracting broad attention.
Surveys frame tasks that cross a context window or session and call for durable history and recovery outside the model~\cite{202607.1328}.
Other work measures growing time spans on software tasks~\cite{kwa2025measuring} and stresses ultra-long tool use~\cite{luo2026ultrahorizon}.
A large body of benchmarks targets long-horizon software-engineering tasks~\cite{deng2026swebench,anonymous2026sweevo,xu2026roadmapbenchevaluatinglonghorizonagentic,ding2026nlrepobench,zhao2025commit,desai2026swemarathonagentsautonomouslycomplete}, long terminal workflows~\cite{li2026longhorizonterminalbenchtestinglimitsagents,feng2026longclibenchpreliminarybenchmarkstudy,merrill2026terminalbench}, research and ML pipelines~\cite{chan2025mlebenchevaluatingmachinelearning,pmlr-v267-starace25a,wijk2025rebench,lu2024aiscientistfullyautomated,yamada2025aiscientistv2workshoplevelautomated,mitchener2025kosmosaiscientistautonomous,rank2026posttrainbenchllmagentsautomate}, and office or web settings~\cite{NEURIPS2025_0d744742,anonymous2026clawmark,li2026the,zhu2026workflowgymlonghorizonevaluationcomputeruse,yoran2024assistantbenchwebagentssolve,zhou2024webarenarealisticwebenvironment,3737916.3739566,he2026ycbench}.
These efforts evaluate long-horizon capability.
\sys{} complements them by representing and reusing the project context such work depends on.

\myparagraph{Content addressing and functional packages.}
Prior systems identify each block or file by a hash of its contents, so identical data is kept once and older snapshots stay readable~\cite{270750,1404014.1404025,10.1145/502059.502054,271167,1267359.1267365,10.1145/502059.502052}.
Git~\cite{chacon2014gitobjects}, Ori~\cite{10.1145/2517349.2522721}, Software Heritage~\cite{10.1145/3379597.3387510}, IPFS~\cite{10.1145/3544216.3544232}, and OSTree~\cite{ostree} use the same idea for directory trees and history.
Nix~\cite{269552}, NixOS~\cite{10.1145/1411204.1411255}, and Guix~\cite{courtès2013functionalpackagemanagementguix} put each package under a path fixed by its build inputs, so many versions can sit side by side and an upgrade only switches which one is active.
\sys{} builds on these ideas for agent project memory, where generation switching selects which context files a project exposes to later tasks.

\section{Discussion}

\myparagraph{Workload generality.}
Precomputing what a project should read works when that set stays fairly steady.
It is a weaker fit when the selection keeps moving,
for example
in open-ended search across repositories,
or with trajectories that change every turn.
Those cases still need runtime lookup, and \sys{} leaves that path open.
Its memory tree is readable with ordinary file tools,
so an agent can browse and search global memory,
or look up the intent text attached to a Quest,
without first binding anything into the project.
The same objects can also carry embeddings or summaries,
so familiar retrieval methods can integrate with \sys{}.

\myparagraph{Remote stores and shared memory.}
The prototype we evaluate keeps memory next to the project on one machine,
but a team often wants one shared Store and many local checkouts.
\sys{} can use a version control system (e.g., Git~\cite{git-system}) to synchronize the Store across machines.
Because each file is addressed by content and each view records the hashes of the files it depends on,
a checkout can fetch only the sources it needs and build the View locally.
Objects are immutable,
so content from two machines will never conflict.

\myparagraph{Deletion and retention.}
\sys{} does not reclaim Store objects automatically when no view currently points at them.
Package stores can do that because an unused closure is safe to drop; agent memory is different.
A file with no live view today may still be needed in the future,
for example when a later debug session restores an earlier project binding.
So the Store keeps history by default, and removal is an explicit user action.

\myparagraph{View selection.}
Binding a project to a View is itself a choice,
and the agent can make mistakes.
For example,
two Views with similar intents can both look right for a task.
We argue that this choice is easier to get right than retrieve-at-use.
First,
it is made over a small set of intents,
and the agent can read the Store files behind each candidate View before binding anything.
Second,
the binding is explicit and recorded,
so a wrong choice is visible afterwards,
and the record turns the mistake into feedback the agent can act on.
Third,
correction is cheap.
Switching the binding moves the whole project to another precomputed view in one step.

\section{Conclusion}
\label{s:conclusion}

We present \sys,
an agent memory system that precomputes context selection as a first-class object with stable identity.
Guided by the insight that what an agent should read is mostly determined by its project,
\sys replaces use-time assembly with
a versioned \emph{Store},
composable \emph{Views},
and \emph{Quests} and \emph{Profiles} that refresh and bind the delivered View.
Hence,
this work opens new directions for building shared, durable memory ecosystems for agents.

\bibliographystyle{ACM-Reference-Format} %
\bibliography{ref}

@online{git-system,
  author       = {{Git Project}},
  year         = {2026},
  title        = {{Git}},
  url          = {https://git-scm.com/},
  lastaccessed = {April 16, 2025}
}

@online{codex,
  author       = {{OpenAI}},
  year         = {2025},
  title        = {{Introducing Codex}},
  url          = {https://openai.com/index/introducing-codex/},
  lastaccessed = {May 16, 2025}
}

@online{claudecode,
  author       = {{Anthropic}},
  year         = {2026},
  title        = {{Claude Code Overview}},
  url          = {https://code.claude.com/docs/en/overview},
  lastaccessed = {May 14, 2026}
}

@online{pi,
  author       = {{Pi}},
  year         = {2026},
  title        = {{Pi Coding Agent}},
  url          = {https://pi.dev/},
  lastaccessed = {August 24, 2026}
}

@misc{robbes2026agenticmuchadoptioncoding,
  title         = {Agentic Much? Adoption of Coding Agents on GitHub},
  author        = {Romain Robbes and Théo Matricon and Thomas Degueule and Andre Hora
                   and Stefano Zacchiroli},
  year          = {2026},
  eprint        = {2601.18341},
  archiveprefix = {arXiv},
  primaryclass  = {cs.SE},
  url           = {https://arxiv.org/abs/2601.18341}
}

@misc{watanabe2026useagenticcodingempirical,
  title         = {On the Use of Agentic Coding: An Empirical Study of Pull Requests
                   on GitHub},
  author        = {Miku Watanabe and Hao Li and Yutaro Kashiwa and Brittany Reid and
                   Hajimu Iida and Ahmed E. Hassan},
  year          = {2026},
  eprint        = {2509.14745},
  archiveprefix = {arXiv},
  primaryclass  = {cs.SE},
  url           = {https://arxiv.org/abs/2509.14745}
}

@inproceedings{Li_2026,
  series     = {MSR ’26},
  title      = {AIDev: Studying AI Coding Agents on GitHub},
  url        = {http://dx.doi.org/10.1145/3793302.3797249},
  doi        = {10.1145/3793302.3797249},
  booktitle  = {Proceedings of the 23rd International Conference on Mining
                Software Repositories},
  publisher  = {Association for Computing Machinery},
  address    = {New York, NY, USA},
  author     = {Li, Hao and Zhang, Haoxiang and Hassan, Ahmed E.},
  year       = {2026},
  month      = Apr,
  pages      = {1029--1033},
  collection = {MSR ’26}
}

@online{effective_harness,
  author       = {{Anthropic}},
  year         = {2025},
  title        = {{Effective Harnesses for Long-Running Agents}},
  url          = {
                  https://www.anthropic.com/engineering/effective-harnesses-for-long-running-agents
                  },
  lastaccessed = {November 26, 2025}
}

@misc{vasilopoulos2026codifiedcontextinfrastructureai,
  title         = {Codified Context: Infrastructure for AI Agents in a Complex
                   Codebase},
  author        = {Aristidis Vasilopoulos},
  year          = {2026},
  eprint        = {2602.20478},
  archiveprefix = {arXiv},
  primaryclass  = {cs.SE},
  url           = {https://arxiv.org/abs/2602.20478}
}

@misc{kwa2026measuringaiabilitycomplete,
  title         = {Measuring AI Ability to Complete Long Software Tasks},
  author        = {Thomas Kwa and Ben West and Joel Becker and Amy Deng and Katharyn
                   Garcia and Max Hasin and Sami Jawhar and Megan Kinniment and Nate
                   Rush and Sydney Von Arx and Ryan Bloom and Thomas Broadley and
                   Haoxing Du and Brian Goodrich and Nikola Jurkovic and Luke Harold
                   Miles and Seraphina Nix and Tao Lin and Chris Painter and Neev
                   Parikh and David Rein and Lucas Jun Koba Sato and Hjalmar Wijk and
                   Daniel M. Ziegler and Elizabeth Barnes and Lawrence Chan},
  year          = {2026},
  eprint        = {2503.14499},
  archiveprefix = {arXiv},
  primaryclass  = {cs.AI},
  url           = {https://arxiv.org/abs/2503.14499}
}

@online{claude_cowork,
  author       = {{Anthropic}},
  year         = {2026},
  title        = {{Claude Cowork}},
  url          = {https://claude.com/product/cowork},
  lastaccessed = {August 26, 2026}
}

@online{workbuddy,
  author       = {{WorkBuddy}},
  year         = {2026},
  title        = {{WorkBuddy}},
  url          = {https://www.workbuddy.ai/},
  lastaccessed = {August 26, 2026}
}

@online{openclaw,
  author       = {{OpenClaw}},
  year         = {2026},
  title        = {{OpenClaw}},
  url          = {https://openclaw.ai/},
  lastaccessed = {January 29, 2026}
}

@online{hermes,
  author       = {{Nous Research}},
  year         = {2026},
  title        = {{Hermes Agent}},
  url          = {https://hermes-agent.nousresearch.com/},
  lastaccessed = {August 26, 2026}
}

@online{skill,
  author       = {{Agent Skills}},
  year         = {2026},
  title        = {{Agent Skills Overview}},
  url          = {https://agentskills.io/home},
  lastaccessed = {September 10, 2026}
}

@online{openai2026buildskills,
  author       = {{OpenAI}},
  year         = {2026},
  title        = {{Build Skills}},
  url          = {https://learn.chatgpt.com/docs/build-skills},
  lastaccessed = {September 21, 2026}
}

@online{anthropic2026skills,
  author       = {{Anthropic}},
  year         = {2026},
  title        = {{Extend Claude with Skills}},
  url          = {https://code.claude.com/docs/en/skills},
  lastaccessed = {September 21, 2026}
}

@book{chacon2014gitobjects,
  author    = {Chacon, Scott and Straub, Ben},
  title     = {Pro {Git}},
  year      = {2014},
  edition   = {2nd},
  publisher = {Apress},
  address   = {Berkeley, CA, USA},
  doi       = {10.1007/978-1-4842-0076-6}
}

@inproceedings{10.1109/MSR.2019.00030,
  author    = {Pietri, Antoine and Spinellis, Diomidis and Zacchiroli, Stefano},
  title     = {The software heritage graph dataset: public software development
               under one roof},
  year      = {2019},
  publisher = {IEEE Press},
  address   = {Piscataway, NJ, USA},
  url       = {https://doi.org/10.1109/MSR.2019.00030},
  doi       = {10.1109/MSR.2019.00030},
  booktitle = {Proceedings of the 16th International Conference on Mining
               Software Repositories},
  pages     = {138--142},
  numpages  = {5},
  location  = {Montreal, Quebec, Canada},
  series    = {MSR '19}
}

@online{readthedocsVersions,
  author       = {{Read the Docs}},
  year         = {2026},
  title        = {{Versions}},
  url          = {https://docs.readthedocs.com/platform/latest/versions.html},
  lastaccessed = {September 14, 2026}
}

@inproceedings{low2023git,
  author    = {Yucheng Low and Rajat Arya and Ajit Banerjee and Ann Huang and
               Brian Ronan and Hoyt Koepke and Joseph Godlewski and Zach Nation},
  title     = {{Git} Is for Data},
  booktitle = {Proceedings of the 13th Annual Conference on Innovative Data
               Systems Research ({CIDR} '23)},
  year      = {2023},
  month     = jan,
  address   = {Amsterdam, The Netherlands},
  publisher = {www.cidrdb.org},
  numpages  = {9},
  url       = {https://www.vldb.org/cidrdb/papers/2023/p43-low.pdf}
}

@misc{li2025memos,
  author        = {Zhiyu Li and Chenyang Xi and Chunyu Li and Ding Chen and Boyu Chen
                   and Shichao Song and Simin Niu and Hanyu Wang and Jiawei Yang and
                   Chen Tang and Qingchen Yu and Jihao Zhao and Yezhaohui Wang and
                   Peng Liu and Zehao Lin and Pengyuan Wang and Jiahao Huo and Tianyi
                   Chen and Kai Chen and Kehang Li and Zhen Tao and Huayi Lai and Hao
                   Wu and Bo Tang and Zhengren Wang and Zhaoxin Fan and Ningyu Zhang
                   and Linfeng Zhang and Junchi Yan and Mingchuan Yang and Tong Xu and
                   Wei Xu and Huajun Chen and Haofen Wang and Hongkang Yang and Wentao
                   Zhang and Zhi-Qin John Xu and Siheng Chen and Feiyu Xiong},
  title         = {{MemOS}: A Memory {OS} for {AI} System},
  year          = {2025},
  eprint        = {2507.03724},
  archiveprefix = {arXiv},
  primaryclass  = {cs.CL},
  doi           = {10.48550/arXiv.2507.03724},
  url           = {https://arxiv.org/abs/2507.03724}
}

@inproceedings{10.1145/3448016.3457550,
  author    = {Wang, Jianguo and Yi, Xiaomeng and Guo, Rentong and Jin, Hai and
               Xu, Peng and Li, Shengjun and Wang, Xiangyu and Guo, Xiangzhou and
               Li, Chengming and Xu, Xiaohai and Yu, Kun and Yuan, Yuxing and Zou,
               Yinghao and Long, Jiquan and Cai, Yudong and Li, Zhenxiang and
               Zhang, Zhifeng and Mo, Yihua and Gu, Jun and Jiang, Ruiyi and Wei,
               Yi and Xie, Charles},
  title     = {Milvus: A Purpose-Built Vector Data Management System},
  year      = {2021},
  isbn      = {9781450383431},
  publisher = {Association for Computing Machinery},
  address   = {New York, NY, USA},
  url       = {https://doi.org/10.1145/3448016.3457550},
  doi       = {10.1145/3448016.3457550},
  booktitle = {Proceedings of the 2021 International Conference on Management
               of Data},
  pages     = {2614--2627},
  numpages  = {14},
  location  = {Virtual Event, China},
  series    = {SIGMOD '21}
}

@inproceedings{DBLP:conf/ecai/ChhikaraKASY25,
  author    = {Prateek Chhikara and Dev Khant and Saket Aryan and Taranjeet Singh
               and Deshraj Yadav},
  editor    = {In{\^{e}}s Lynce and Nello Murano and Mauro Vallati and Serena
               Villata and Federico Chesani and Michela Milano and Andrea Omicini
               and Mehdi Dastani},
  title     = {Mem0: Building Production-Ready {AI} Agents with Scalable Long-Term
               Memory},
  booktitle = {{ECAI} 2025 - 28th European Conference on Artificial
               Intelligence, 25-30 October 2025, Bologna, Italy - Including
               14th Conference on Prestigious Applications of Intelligent
               Systems {(PAIS} 2025)},
  series    = {Frontiers in Artificial Intelligence and Applications},
  volume    = {413},
  pages     = {2993--3000},
  publisher = {{IOS} Press},
  address   = {Amsterdam, The Netherlands},
  year      = {2025},
  url       = {https://doi.org/10.3233/FAIA251160},
  doi       = {10.3233/FAIA251160},
  bibsource = {dblp computer science bibliography, https://dblp.org}
}

@online{bytedance2026openviking,
  author       = {{ByteDance}},
  year         = {2026},
  title        = {{{OpenViking}: The Context Database for {AI} Agents}},
  url          = {https://github.com/volcengine/OpenViking},
  lastaccessed = {September 14, 2026}
}

@article{10.14778/3827998.3828037,
  author     = {Fu, Jiajie and Chen, Junwen and Wang, Mengzhao and He, Aoxiang and
                Sheng, Maojia and Ke, Xiangyu and Zhu, Yifan and Gao, Yunjun},
  title      = {VikingMem: A Memory Base Management System for Stateful LLM-Based
                Applications},
  year       = {2026},
  issue_date = {August 2026},
  publisher  = {VLDB Endowment},
  volume     = {19},
  number     = {12},
  issn       = {2150-8097},
  url        = {https://doi.org/10.14778/3827998.3828037},
  doi        = {10.14778/3827998.3828037},
  journal    = {Proc. VLDB Endow.},
  month      = sep,
  pages      = {4344--4357},
  numpages   = {14}
}

@inproceedings{3495724.3496517,
  author    = {Lewis, Patrick and Perez, Ethan and Piktus, Aleksandra and Petroni
               , Fabio and Karpukhin, Vladimir and Goyal, Naman and K{\"u}ttler,
               Heinrich and Lewis, Mike and Yih, Wen-tau and Rockt{\"a}schel, Tim
               and Riedel, Sebastian and Kiela, Douwe},
  title     = {Retrieval-augmented generation for knowledge-intensive NLP tasks},
  year      = {2020},
  isbn      = {9781713829546},
  publisher = {Curran Associates Inc.},
  address   = {Red Hook, NY, USA},
  booktitle = {Proceedings of the 34th International Conference on Neural
               Information Processing Systems},
  articleno = {793},
  numpages  = {16},
  location  = {Vancouver, BC, Canada},
  series    = {NIPS '20}
}

@inproceedings{karpukhin-etal-2020-dense,
  title     = {Dense Passage Retrieval for Open-Domain Question Answering},
  author    = {Karpukhin, Vladimir and Oguz, Barlas and Min, Sewon and Lewis,
               Patrick and Wu, Ledell and Edunov, Sergey and Chen, Danqi and Yih,
               Wen-tau},
  editor    = {Webber, Bonnie and Cohn, Trevor and He, Yulan and Liu, Yang},
  booktitle = {Proceedings of the 2020 Conference on Empirical Methods in
               Natural Language Processing (EMNLP)},
  month     = nov,
  year      = {2020},
  address   = {Online},
  publisher = {Association for Computational Linguistics},
  url       = {https://aclanthology.org/2020.emnlp-main.550/},
  doi       = {10.18653/v1/2020.emnlp-main.550},
  pages     = {6769--6781}
}

@inproceedings{10.1145/3586183.3606763,
  author    = {Park, Joon Sung and O'Brien, Joseph and Cai, Carrie Jun and Morris
               , Meredith Ringel and Liang, Percy and Bernstein, Michael S.},
  title     = {Generative Agents: Interactive Simulacra of Human Behavior},
  year      = {2023},
  isbn      = {9798400701320},
  publisher = {Association for Computing Machinery},
  address   = {New York, NY, USA},
  url       = {https://doi.org/10.1145/3586183.3606763},
  doi       = {10.1145/3586183.3606763},
  booktitle = {Proceedings of the 36th Annual ACM Symposium on User Interface
               Software and Technology},
  articleno = {2},
  numpages  = {22},
  location  = {San Francisco, CA, USA},
  series    = {UIST '23}
}

@misc{packer2023memgpt,
  title         = {{MemGPT}: Towards LLMs as Operating Systems},
  author        = {Packer, Charles and Wooders, Sarah and Lin, Kevin and Fang, Vivian
                   and Patil, Shishir G. and Stoica, Ion and Gonzalez, Joseph E.},
  year          = {2023},
  eprint        = {2310.08560},
  archiveprefix = {arXiv},
  primaryclass  = {cs.AI}
}

@inproceedings{ICLR2024_8a2acd17,
  author    = {Sarthi, Parth and Abdullah, Salman and Tuli, Aditi and Khanna,
               Shubh and Goldie, Anna and Manning, Christopher},
  booktitle = {International Conference on Learning Representations},
  editor    = {B. Kim and Y. Yue and S. Chaudhuri and K. Fragkiadaki and M. Khan
               and Y. Sun},
  pages     = {32628--32649},
  title     = {RAPTOR: Recursive Abstractive Processing for Tree-Organized
               Retrieval},
  url       = {
               https://proceedings.iclr.cc/paper_files/paper/2024/file/8a2acd174940dbca361a6398a4f9df91-Paper-Conference.pdf
               },
  volume    = {2024},
  year      = {2024},
  publisher = {ICLR},
  address   = {Vienna, Austria}
}

@online{openai2026agentsmd,
  author       = {{OpenAI}},
  year         = {2026},
  title        = {{Custom Instructions with {AGENTS.md}}},
  url          = {https://learn.chatgpt.com/docs/agent-configuration/agents-md},
  lastaccessed = {September 14, 2026}
}

@online{anthropic2026claudememory,
  author       = {{Anthropic}},
  year         = {2026},
  title        = {{How {Claude} Remembers Your Project}},
  url          = {https://code.claude.com/docs/en/memory},
  lastaccessed = {September 14, 2026}
}

@online{langchain2026langgraphpersistence,
  author       = {{LangChain}},
  year         = {2026},
  title        = {{{LangGraph} Persistence}},
  url          = {https://docs.langchain.com/oss/python/langgraph/persistence},
  lastaccessed = {September 14, 2026}
}

@inproceedings{ICLR2025_a4b6ad6b,
  author    = {Wang, Xingyao and Li, Boxuan and Song, Yufan and Xu, Frank F and
               Tang, Xiangru and Zhuge, Mingchen and Pan, Jiayi and Song, Yueqi
               and Li, Bowen and Singh, Jaskirat and Tran, Hoang and Li, Fuqiang
               and Ma, Ren and Zheng, Mingzhang and Qian, Bill and Shao, Daniel
               and Muennighoff, Niklas and Zhang, Yizhe and Hui, Binyuan and Lin,
               Junyang and Brennan, Robert and Peng, Hao and Ji, Heng and Neubig,
               Graham},
  booktitle = {International Conference on Learning Representations},
  editor    = {Y. Yue and A. Garg and N. Peng and F. Sha and R. Yu},
  pages     = {65882--65919},
  title     = {OpenHands: An Open Platform for AI Software Developers as
               Generalist Agents},
  url       = {
               https://proceedings.iclr.cc/paper_files/paper/2025/file/a4b6ad6b48850c0c331d1259fc66a69c-Paper-Conference.pdf
               },
  volume    = {2025},
  year      = {2025},
  publisher = {ICLR},
  address   = {Singapore}
}

@misc{dong2024agentopsenablingobservabilityllm,
  title         = {AgentOps: Enabling Observability of LLM Agents},
  author        = {Liming Dong and Qinghua Lu and Liming Zhu},
  year          = {2024},
  eprint        = {2411.05285},
  archiveprefix = {arXiv},
  primaryclass  = {cs.AI},
  url           = {https://arxiv.org/abs/2411.05285}
}

@inproceedings{270438,
  author    = {George W. Dunlap and Samuel T. King and Sukru Cinar and Murtaza A.
               Basrai and Peter M. Chen},
  title     = {{ReVirt}: Enabling Intrusion Analysis Through {Virtual-Machine}
               Logging and Replay},
  booktitle = {5th Symposium on Operating Systems Design and Implementation
               (OSDI 02)},
  year      = {2002},
  address   = {Boston, MA},
  pages     = {211--224},
  url       = {
               https://www.usenix.org/conference/osdi-02/revirt-enabling-intrusion-analysis-through-virtual-machine-logging-and-replay
               },
  publisher = {USENIX Association},
  month     = dec
}

@inproceedings{203227,
  author    = {Robert O{\textquoteright}Callahan and Chris Jones and Nathan Froyd
               and Kyle Huey and Albert Noll and Nimrod Partush},
  title     = {Engineering Record and Replay for Deployability},
  booktitle = {2017 USENIX Annual Technical Conference (USENIX ATC 17)},
  year      = {2017},
  isbn      = {978-1-931971-38-6},
  address   = {Santa Clara, CA},
  pages     = {377--389},
  url       = {
               https://www.usenix.org/conference/atc17/technical-sessions/presentation/ocallahan
               },
  publisher = {USENIX Association},
  month     = jul
}

@inproceedings{Ansel_2009,
  author    = {Ansel, Jason and Arya, Kapil and Cooperman, Gene},
  title     = {{DMTCP}: Transparent Checkpointing for Cluster Computations and the
               Desktop},
  booktitle = {2009 IEEE International Symposium on Parallel \& Distributed
               Processing},
  year      = {2009},
  month     = may,
  pages     = {1--12},
  publisher = {IEEE},
  address   = {Piscataway, NJ, USA},
  doi       = {10.1109/IPDPS.2009.5161063},
  url       = {https://doi.org/10.1109/IPDPS.2009.5161063}
}

@misc{dong2026deltaboxscalingstatefulai,
  title         = {DeltaBox: Scaling Stateful AI Agents with Millisecond-Level Sandbox
                   Checkpoint/Rollback},
  author        = {Yunpeng Dong and Jingkai He and Shiqi Liu and Yuze Hou and Dong Du
                   and Zhonghu Xu and Si Yu and Baochuan Yang and Yubin Xia and Haibo
                   Chen},
  year          = {2026},
  eprint        = {2605.22781},
  archiveprefix = {arXiv},
  primaryclass  = {cs.OS},
  url           = {https://arxiv.org/abs/2605.22781}
}

@online{google2026geminimd,
  author       = {{Google}},
  year         = {2026},
  title        = {{Provide Context with {GEMINI.md} Files}},
  url          = {https://geminicli.com/docs/cli/gemini-md/},
  lastaccessed = {September 15, 2026}
}

@online{cursor2026rules,
  author       = {{Cursor}},
  year         = {2026},
  title        = {{Rules}},
  url          = {https://cursor.com/docs/rules},
  lastaccessed = {September 15, 2026}
}

@misc{zhu2026swecontextbenchbenchmark,
  title         = {SWE Context Bench: A Benchmark for Context Learning in Coding},
  author        = {Jiayuan Zhu and Junde Wu and Minhao Hu and Shengda Zhu and Jiazhen
                   Pan and Weixiang Shen and Yijun Yang and Fenglin Liu and Jianye Hao
                   and Yueming Jin and Qirong Ho and Min Xu},
  year          = {2026},
  eprint        = {2602.08316},
  archiveprefix = {arXiv},
  primaryclass  = {cs.SE},
  url           = {https://arxiv.org/abs/2602.08316}
}

@online{mem0-open-source-overview,
  author       = {{Mem0}},
  year         = {2026},
  title        = {{Open Source Overview}},
  url          = {https://docs.mem0.ai/open-source/overview},
  lastaccessed = {September 17, 2026}
}

@online{deepseek-api-pricing-2026,
  author       = {{DeepSeek}},
  year         = {2026},
  title        = {Models \& Pricing},
  url          = {https://api-docs.deepseek.com/quick_start/pricing/},
  lastaccessed = {September 17, 2026}
}

@techreport{deepseek-v41-technical-report-2026,
  author       = {{DeepSeek-AI}},
  year         = {2026},
  title        = {{DeepSeek-V4.1-Flash}: Pushing the Limits of {KV} Cache Compression
                  },
  institution  = {DeepSeek-AI},
  url          = {
                  https://huggingface.co/deepseek-ai/DeepSeek-V4.1-Flash/blob/main/DeepSeek_V41_Tech_Report.pdf
                  },
  lastaccessed = {September 17, 2026}
}

@online{openai-text-embedding-3-large,
  author       = {{OpenAI}},
  year         = {2026},
  title        = {{text-embedding-3-large} Model},
  url          = {https://developers.openai.com/api/docs/models/text-embedding-3-large},
  lastaccessed = {September 17, 2026}
}

@inproceedings{kuhar-etal-2025-libevolutioneval,
  title     = {{L}ib{E}volution{E}val: A Benchmark and Study for Version-Specific
               Code Generation},
  author    = {Kuhar, Sachit and Ahmad, Wasi Uddin and Wang, Zijian and Jain,
               Nihal and Qian, Haifeng and Ray, Baishakhi and Ramanathan, Murali
               Krishna and Ma, Xiaofei and Deoras, Anoop},
  editor    = {Chiruzzo, Luis and Ritter, Alan and Wang, Lu},
  booktitle = {Proceedings of the 2025 Conference of the Nations of the
               Americas Chapter of the Association for Computational
               Linguistics: Human Language Technologies (Volume 1: Long Papers)
               },
  month     = apr,
  year      = {2025},
  address   = {Albuquerque, New Mexico},
  publisher = {Association for Computational Linguistics},
  url       = {https://aclanthology.org/2025.naacl-long.348/},
  doi       = {10.18653/v1/2025.naacl-long.348},
  pages     = {6826--6840},
  isbn      = {979-8-89176-189-6}
}

@article{202607.1328,
  doi       = {10.20944/preprints202607.1328.v1},
  url       = {https://doi.org/10.20944/preprints202607.1328.v1},
  year      = 2026,
  month     = {July},
  publisher = {Preprints},
  author    = {Guanting Dong and Xiaoshuai Song and Yuyang Hu and Jiajie Jin and
               Chenghao Zhang and Yifei Chen and Xiaoxi Li and Huaying Yuan and
               Xinyu Yang and Tongyu Wen and Jiejun Tan and Hongjin Qian and
               Shijue Huang and Junting Lu and Zhenyu Li and Wanjun Zhong and
               Yutao Zhu and Tat-Seng Chua and Zhicheng Dou and Ji-Rong Wen},
  title     = {Towards Long-Horizon Agents: A Survey},
  journal   = {Preprints}
}

@inproceedings{kwa2025measuring,
  title     = {Measuring {AI} Ability to Complete Long Software Tasks},
  author    = {Thomas Kwa and Ben West and Joel Becker and Amy Deng and Katharyn
               Garcia and Max Hasin and Sami Jawhar and Megan Kinniment and Nate
               Rush and Sydney Von Arx and Ryan Bloom and Thomas Broadley and
               Haoxing Du and Brian Goodrich and Nikola Jurkovic and Luke Harold
               Miles and Seraphina Nix and Tao Roa Lin and Neev Parikh and David
               Rein and Lucas Jun Koba Sato and Hjalmar Wijk and Daniel M Ziegler
               and Elizabeth Barnes and Lawrence Chan},
  booktitle = {The Thirty-ninth Annual Conference on Neural Information
               Processing Systems},
  year      = {2025},
  url       = {https://openreview.net/forum?id=CGNJL6CeV0}
}

@misc{luo2026ultrahorizon,
  title  = {UltraHorizon: Benchmarking {LLM}-Agent Capabilities in Ultra
            Long-Horizon Scenarios},
  author = {Haotian Luo and Huaisong Zhang and Xuelin Zhang and Haoyu Wang and
            Zeyu Qin and WenJie Lu and Guozheng Ma and Haiying He and Yingsha
            Xie and Qiyang Zhou and Zixuan Hu and Hongze Mi and Yibo Wang and
            Naiqiang Tan and Hong Chen and Yi R. Fung and Chun Yuan and Li Shen
            },
  year   = {2026},
  url    = {https://openreview.net/forum?id=FTZfVHWAIq}
}

@inproceedings{deng2026swebench,
  title     = {{SWE}-Bench Pro: Can {AI} Agents Solve Long-Horizon Software
               Engineering Tasks?},
  author    = {Xiang Deng and Jeff Da and Edwin Pan and Yannis Yiming He and
               Charles Ide and Kanak Garg and Niklas Lauffer and Andrew Park and
               Chetan Rane and Karmini Sampath and Maya Krishnan and Srivatsa R
               Kundurthy and Sean M. Hendryx and Zifan Wang and Chen Bo Calvin
               Zhang and Noah Jacobson and Bing Liu and Brad Kenstler},
  booktitle = {Forty-third International Conference on Machine Learning},
  year      = {2026},
  url       = {https://openreview.net/forum?id=uEVTdoAbnK}
}

@inproceedings{anonymous2026sweevo,
  title     = {{SWE}-{EVO}: Benchmarking Coding Agents in Long-Horizon Software
               Evolution Scenarios},
  author    = {Anonymous},
  booktitle = {Submitted to ACL Rolling Review - March 2026},
  year      = {2026},
  url       = {https://openreview.net/forum?id=SCpCfeSLtn},
  note      = {under review}
}

@misc{xu2026roadmapbenchevaluatinglonghorizonagentic,
  title         = {RoadmapBench: Evaluating Long-Horizon Agentic Software Development
                   Across Version Upgrades},
  author        = {Xinbo Xu and Ruihan Yang and Haiyang Shen and Wendong Xu and Bofei
                   Gao and Ruoyu Wu and Kean Shi and Weichu Xie and Xuanzhong Chen and
                   Ming Wu and Jason Zeng and Michael Heinrich and Elvis Zhang and
                   Liang Chen and Kuan Li and Baobao Chang},
  year          = {2026},
  eprint        = {2605.15846},
  archiveprefix = {arXiv},
  primaryclass  = {cs.SE},
  url           = {https://arxiv.org/abs/2605.15846}
}

@inproceedings{ding2026nlrepobench,
  title     = {{NL}2Repo-Bench: Towards Long-Horizon Repository Generation
               Evaluation of Coding Agents},
  author    = {Jingzhe Ding and Shengda Long and Changxin PU and Ge Zhang and
               zhou huan and Hongwan Gao and Xiang Gao and Chao He and Yue Hou and
               FEI HU and Zhaojian Li and Weiran Shi and Zaiyuan Wang and Daoguang
               Zan and Chenchen Zhang and Xiaoxu Zhang and Chen Qizhi and Xianfu
               Cheng and Bo Deng and Qingshui Gu and Kai Hua and Juntao Lin and
               Pai Liu and Mingchen Li and Minghao Li and Xuanguang Pan and Zifan
               Peng and Yujia Qin and Yong Shan and Zhewen Tan and Haoran Wang and
               Zihan Wang and Weihao Xie and Yishuo Yuan and Jiayu Zhang and
               Yunfei Zhao and He Zhu and LIYA ZHU and Chenyang Zou and Ming Ding
               and Jianpeng Jiao and Jiaheng Liu and Minghao Liu and Qian Liu and
               Chongyang Tao and Jian Yang and Tong Yang and Zhaoxiang Zhang and
               Xinjie Chen and Wenhao Huang},
  booktitle = {Forty-third International Conference on Machine Learning},
  year      = {2026},
  url       = {https://openreview.net/forum?id=wqQam1muOQ}
}

@inproceedings{zhao2025commit,
  title     = {Commit0: Library Generation from Scratch},
  author    = {Wenting Zhao and Nan Jiang and Celine Lee and Justin T Chiu and
               Claire Cardie and Matthias Gall{\'e} and Alexander M Rush},
  booktitle = {The Thirteenth International Conference on Learning
               Representations},
  year      = {2025},
  url       = {https://openreview.net/forum?id=MMwaQEVsAg}
}

@misc{desai2026swemarathonagentsautonomouslycomplete,
  title         = {SWE-Marathon: Can Agents Autonomously Complete Ultra-Long-Horizon
                   Software Work?},
  author        = {Rishi Desai and Jesse Hu and Joan Cabezas and Neel Harsola and
                   Pratyush Shukla and Roey Ben Chaim and Adnan El Assadi and Omkaar
                   Mukund Kamath and Fenil Faldu and Prannay Hebbar and Jiankai Sun
                   and Yiyuan Li and Pramod Srinivasan and Ishan Gupta and Christopher
                   Settles and Daniel Wang and Derek Chen and Pranav Raja and Albert
                   Liu and Marek Šuppa and Nevasini Sasikumar and Luyang Kong and Erik
                   Quintanilla and Xiangyi Li and Ivan Bercovich and Steven Dillmann},
  year          = {2026},
  eprint        = {2606.07682},
  archiveprefix = {arXiv},
  primaryclass  = {cs.SE},
  url           = {https://arxiv.org/abs/2606.07682}
}

@misc{li2026longhorizonterminalbenchtestinglimitsagents,
  title         = {Long-Horizon-Terminal-Bench: Testing the Limits of Agents on
                   Long-Horizon Terminal Tasks with Dense Reward-Based Grading},
  author        = {Zongxia Li and Zhongzhi Li and Yucheng Shi and Ruhan Wang and
                   Junyao Yang and Zhichao Liu and Xiyang Wu and Anhao Li and Yue Yu
                   and Ninghao Liu and Lichao Sun and Haotao Mi and Leowei Liang},
  year          = {2026},
  eprint        = {2607.08964},
  archiveprefix = {arXiv},
  primaryclass  = {cs.AI},
  url           = {https://arxiv.org/abs/2607.08964}
}

@misc{feng2026longclibenchpreliminarybenchmarkstudy,
  title         = {LongCLI-Bench: A Preliminary Benchmark and Study for Long-horizon
                   Agentic Programming in Command-Line Interfaces},
  author        = {Yukang Feng and Jianwen Sun and Zelai Yang and Jiaxin Ai and
                   Chuanhao Li and Zizhen Li and Fanrui Zhang and Kang He and Rui Ma
                   and Jifan Lin and Jie Sun and Yang Xiao and Sizhuo Zhou and Wenxiao
                   Wu and Yiming Liu and Pengfei Liu and Yu Qiao and Shenglin Zhang
                   and Kaipeng Zhang},
  year          = {2026},
  eprint        = {2602.14337},
  archiveprefix = {arXiv},
  primaryclass  = {cs.SE},
  url           = {https://arxiv.org/abs/2602.14337}
}

@inproceedings{merrill2026terminalbench,
  title     = {Terminal-Bench: Benchmarking Agents on Hard, Realistic Tasks in
               Command Line Interfaces},
  author    = {Mike A Merrill and Alexander Glenn Shaw and Nicholas Carlini and
               Boxuan Li and Harsh Raj and Ivan Bercovich and Lin Shi and Jeong
               Yeon Shin and Thomas Walshe and E. Kelly Buchanan and Junhong Shen
               and Guanghao Ye and Haowei Lin and Jason Poulos and Maoyu Wang and
               Marianna Nezhurina and Di Lu and Orfeas Menis Mastromichalakis and
               Zhiwei Xu and Zizhao Chen and Yue Liu and Robert Zhang and Leon
               Liangyu Chen and Anurag Kashyap and Jan-Lucas Uslu and Jeffrey Li
               and Jianbo Wu and Minghao Yan and Song Bian and Vedang Sharma and
               Ke Sun and Steven Dillmann and Akshay Anand and Andrew
               Lanpouthakoun and Bardia Koopah and Changran Hu and Etash Kumar
               Guha and Gabriel H. S. Dreiman and Jiacheng Zhu and Karl Krauth and
               Li Zhong and Niklas Muennighoff and Robert Kwesi Amanfu and
               Shangyin Tan and Shreyas Pimpalgaonkar and Tushar Aggarwal and
               Xiangning Lin and Xin Lan and Xuandong Zhao and Yiqing Liang and
               Yuanli Wang and Zilong Wang and Changzhi Zhou and David Heineman
               and Hange Liu and Harsh Trivedi and John Yang and Junhong Lin and
               Manish Shetty and Michael Yang and Nabil Omi and Negin Raoof and
               Shanda Li and Terry Yue Zhuo and Wuwei Lin and Yiwei Dai and Yuxin
               Wang and Wenhao Chai and Shang Zhou and Dariush Wahdany and Ziyu
               She and Jiaming Hu and Zhikang Dong and Yuxuan Zhu and Sasha Cui
               and Ahson Saiyed and Arinbj{\"o}rn Kolbeinsson and Christopher
               Michael Rytting and Ryan Marten and Yixin Wang and Jenia Jitsev and
               Alex Dimakis and Andy Konwinski and Ludwig Schmidt},
  booktitle = {The Fourteenth International Conference on Learning
               Representations},
  year      = {2026},
  url       = {https://openreview.net/forum?id=a7Qa4CcHak}
}

@misc{chan2025mlebenchevaluatingmachinelearning,
  title         = {MLE-bench: Evaluating Machine Learning Agents on Machine Learning
                   Engineering},
  author        = {Jun Shern Chan and Neil Chowdhury and Oliver Jaffe and James Aung
                   and Dane Sherburn and Evan Mays and Giulio Starace and Kevin Liu
                   and Leon Maksin and Tejal Patwardhan and Lilian Weng and Aleksander
                   Mądry},
  year          = {2025},
  eprint        = {2410.07095},
  archiveprefix = {arXiv},
  primaryclass  = {cs.CL},
  url           = {https://arxiv.org/abs/2410.07095}
}

@inproceedings{pmlr-v267-starace25a,
  title     = {{P}aper{B}ench: Evaluating {AI}’s Ability to Replicate {AI}
               Research},
  author    = {Starace, Giulio and Jaffe, Oliver and Sherburn, Dane and Aung,
               James and Chan, Jun Shern and Maksin, Leon and Dias, Rachel and
               Mays, Evan and Kinsella, Benjamin and Thompson, Wyatt and Heidecke,
               Johannes and Glaese, Amelia and Patwardhan, Tejal},
  booktitle = {Proceedings of the 42nd International Conference on Machine
               Learning},
  pages     = {56843--56873},
  year      = {2025},
  editor    = {Singh, Aarti and Fazel, Maryam and Hsu, Daniel and Lacoste-Julien,
               Simon and Berkenkamp, Felix and Maharaj, Tegan and Wagstaff, Kiri
               and Zhu, Jerry},
  volume    = {267},
  series    = {Proceedings of Machine Learning Research},
  month     = {13--19 Jul},
  publisher = {PMLR},
  url       = {https://proceedings.mlr.press/v267/starace25a.html}
}

@inproceedings{wijk2025rebench,
  title     = {{RE}-Bench: Evaluating Frontier {AI} R\&D Capabilities of Language
               Model Agents against Human Experts},
  author    = {Hjalmar Wijk and Tao Roa Lin and Joel Becker and Sami Jawhar and
               Neev Parikh and Thomas Broadley and Lawrence Chan and Michael Chen
               and Joshua M Clymer and Jai Dhyani and Elena Ericheva and Katharyn
               Garcia and Brian Goodrich and Nikola Jurkovic and Megan Kinniment
               and Aron Lajko and Seraphina Nix and Lucas Jun Koba Sato and
               William Saunders and Maksym Taran and Ben West and Elizabeth Barnes
               },
  booktitle = {Forty-second International Conference on Machine Learning},
  year      = {2025},
  url       = {https://openreview.net/forum?id=3rB0bVU6z6}
}

@misc{lu2024aiscientistfullyautomated,
  title         = {The AI Scientist: Towards Fully Automated Open-Ended Scientific
                   Discovery},
  author        = {Chris Lu and Cong Lu and Robert Tjarko Lange and Jakob Foerster
                   and Jeff Clune and David Ha},
  year          = {2024},
  eprint        = {2408.06292},
  archiveprefix = {arXiv},
  primaryclass  = {cs.AI},
  url           = {https://arxiv.org/abs/2408.06292}
}

@misc{yamada2025aiscientistv2workshoplevelautomated,
  title         = {The AI Scientist-v2: Workshop-Level Automated Scientific Discovery
                   via Agentic Tree Search},
  author        = {Yutaro Yamada and Robert Tjarko Lange and Cong Lu and Shengran Hu
                   and Chris Lu and Jakob Foerster and Jeff Clune and David Ha},
  year          = {2025},
  eprint        = {2504.08066},
  archiveprefix = {arXiv},
  primaryclass  = {cs.AI},
  url           = {https://arxiv.org/abs/2504.08066}
}

@misc{mitchener2025kosmosaiscientistautonomous,
  title         = {Kosmos: An AI Scientist for Autonomous Discovery},
  author        = {Ludovico Mitchener and Angela Yiu and Benjamin Chang and Mathieu
                   Bourdenx and Tyler Nadolski and Arvis Sulovari and Eric C.
                   Landsness and Daniel L. Barabasi and Siddharth Narayanan and Nicky
                   Evans and Shriya Reddy and Martha Foiani and Aizad Kamal and Leah
                   P. Shriver and Fang Cao and Asmamaw T. Wassie and Jon M. Laurent
                   and Edwin Melville-Green and Mayk Caldas and Albert Bou and Kaleigh
                   F. Roberts and Sladjana Zagorac and Timothy C. Orr and Miranda E.
                   Orr and Kevin J. Zwezdaryk and Ali E. Ghareeb and Laurie McCoy and
                   Bruna Gomes and Euan A. Ashley and Karen E. Duff and Tonio
                   Buonassisi and Tom Rainforth and Randall J. Bateman and Michael
                   Skarlinski and Samuel G. Rodriques and Michaela M. Hinks and Andrew
                   D. White},
  year          = {2025},
  eprint        = {2511.02824},
  archiveprefix = {arXiv},
  primaryclass  = {cs.AI},
  url           = {https://arxiv.org/abs/2511.02824}
}

@misc{rank2026posttrainbenchllmagentsautomate,
  title         = {PostTrainBench: Can LLM Agents Automate LLM Post-Training?},
  author        = {Ben Rank and Hardik Bhatnagar and Ameya Prabhu and Shira Eisenberg
                   and Karina Nguyen and Matthias Bethge and Maksym Andriushchenko},
  year          = {2026},
  eprint        = {2603.08640},
  archiveprefix = {arXiv},
  primaryclass  = {cs.SE},
  url           = {https://arxiv.org/abs/2603.08640}
}

@inproceedings{NEURIPS2025_0d744742,
  author    = {Xu, Frank (Fangzheng) and Song, Yufan and Li, Boxuan and Tang,
               Yuxuan and Jain, Kritanjali and Bao, Mengxue and Wang, Zora and
               Zhou, Xuhui and Guo, Zhitong and Cao, Murong and Yang, Mingyang and
               Lu, Hao Yang and Martin, Amaad and Su, Zhe and Maben, Leander and
               Mehta, Raj and Chi, Wayne and Jang, Lawrence and Xie, Yiqing and
               Zhou, Shuyan and Neubig, Graham},
  booktitle = {Advances in Neural Information Processing Systems},
  doi       = {10.52202/085713-0315},
  editor    = {D. Belgrave and C. Zhang and H. Lin and R. Pascanu and P. Koniusz
               and M. Ghassemi and N. Chen},
  pages     = {},
  publisher = {Curran Associates, Inc.},
  title     = {TheAgentCompany: Benchmarking LLM Agents on Consequential Real
               World Tasks},
  url       = {
               https://proceedings.neurips.cc/paper_files/paper/2025/file/0d744742f6fac4d1134c019b7cef3c8a-Paper-Datasets_and_Benchmarks_Track.pdf
               },
  volume    = {38, Main Conference},
  year      = {2025}
}

@inproceedings{anonymous2026clawmark,
  title     = {ClawMark: A Living-World Benchmark for Multi-Turn, Multi-Day,
               Multimodal Coworker Agents},
  author    = {Anonymous},
  booktitle = {Submitted to ACL Rolling Review - May 2026},
  year      = {2026},
  url       = {https://openreview.net/forum?id=qe1ml90zuL},
  note      = {under review}
}

@inproceedings{li2026the,
  title     = {The Tool Decathlon: Benchmarking Language Agents for Diverse,
               Realistic, and Long-Horizon Task Execution},
  author    = {Junlong Li and Wenshuo Zhao and Jian Zhao and Weihao Zeng and
               Haoze Wu and Xiaochen Wang and Rui Ge and Yuxuan Cao and Yuzhen
               Huang and Wei Liu and Junteng Liu and Zhaochen Su and Yiyang Guo
               and Fan Zhou and Lueyang Zhang and Juan Michelini and Xingyao Wang
               and Xiang Yue and Shuyan Zhou and Graham Neubig and Junxian He},
  booktitle = {The Fourteenth International Conference on Learning
               Representations},
  year      = {2026},
  url       = {https://openreview.net/forum?id=z53s5p0qhf}
}

@misc{zhu2026workflowgymlonghorizonevaluationcomputeruse,
  title         = {Workflow-GYM: Towards Long-Horizon Evaluation of Computer-use
                   Agentic tasks in Real-World Professional Fields},
  author        = {Liya Zhu and Jingzhe Ding and Jian Zhang and Jianbo Xue and Shihao
                   Liang and Ge Zhang and Yi Zhu and Duju Zeng and Xiang Gao and
                   Qingshui Gu and Mailun Gao and Huimin Che and Yan Zhao and Peiheng
                   Zhou and Haojun Wang and Chaobo Xian and Lili Le and Chi Wu and
                   Yiwei Liu and Shengda Long and Jiale Yang and Fangzhi Xu and Sijin
                   Wu and Haodong Duan and Chao He and Zhaojian Li and Minchao Wang
                   and Huan Zhou and Jiani Hou and Chuqian Yu and Weiran Shi and
                   Hongwan Gao and Jiamin Chen and Guanhong Chen and Tingqin Luo and
                   Kaiyuan Zhang and Zhixin Yao and Qing Hua and Yuhao Jiang and Jin
                   Chen and Pu Chen and Zhenyu Hu and Xingyu Li and Zhengxuan Jiang
                   and Meng Cao and Tianfeng Long and Haozhe Wang and Mingzhang Wang
                   and Yichen Zhang and Yiming Dai and Chenchen Zhang and Jiaying Wang
                   and Xinying Liu and Xingzu Liu and Lingling Zhang and Xinjie Chen
                   and Yujia Qin and Wangchunshu Zhou and Zhiyong Wu and Yang Liu and
                   Jiaheng Liu and Lei Zhang and Shen Yan and Wenhao Huang and Zaiyuan
                   Wang and Xiaolong Chang},
  year          = {2026},
  eprint        = {2606.11042},
  archiveprefix = {arXiv},
  primaryclass  = {cs.AI},
  url           = {https://arxiv.org/abs/2606.11042}
}

@misc{yoran2024assistantbenchwebagentssolve,
  title         = {AssistantBench: Can Web Agents Solve Realistic and Time-Consuming
                   Tasks?},
  author        = {Ori Yoran and Samuel Joseph Amouyal and Chaitanya Malaviya and Ben
                   Bogin and Ofir Press and Jonathan Berant},
  year          = {2024},
  eprint        = {2407.15711},
  archiveprefix = {arXiv},
  primaryclass  = {cs.CL},
  url           = {https://arxiv.org/abs/2407.15711}
}

@misc{zhou2024webarenarealisticwebenvironment,
  title         = {WebArena: A Realistic Web Environment for Building Autonomous
                   Agents},
  author        = {Shuyan Zhou and Frank F. Xu and Hao Zhu and Xuhui Zhou and Robert
                   Lo and Abishek Sridhar and Xianyi Cheng and Tianyue Ou and Yonatan
                   Bisk and Daniel Fried and Uri Alon and Graham Neubig},
  year          = {2024},
  eprint        = {2307.13854},
  archiveprefix = {arXiv},
  primaryclass  = {cs.AI},
  url           = {https://arxiv.org/abs/2307.13854}
}

@inproceedings{3737916.3739566,
  author    = {Xie, Tianbao and Zhang, Danyang and Chen, Jixuan and Li, Xiaochuan
               and Zhao, Siheng and Cao, Ruisheng and Hua, Toh Jing and Cheng,
               Zhoujun and Shin, Dongchan and Lei, Fangyu and Liu, Yitao and Xu,
               Yiheng and Zhou, Shuyan and Savarese, Silvio and Xiong, Caiming and
               Zhong, Victor and Yu, Tao},
  title     = {OSWORLD: benchmarking multimodal agents for open-ended tasks in
               real computer environments},
  year      = {2024},
  isbn      = {9798331314385},
  publisher = {Curran Associates Inc.},
  address   = {Red Hook, NY, USA},
  booktitle = {Proceedings of the 38th International Conference on Neural
               Information Processing Systems},
  articleno = {1650},
  numpages  = {55},
  location  = {Vancouver, BC, Canada},
  series    = {NIPS '24}
}

@inproceedings{he2026ycbench,
  title     = {{YC}-Bench: Benchmarking {AI} Agents for Long-Term Planning and
               Consistent Execution},
  author    = {Muyu He and Vincent Tu and Adit Jain and Anand Kumar and Sachin
               Patro and Soumyadeep Bakshi and Nazneen Rajani},
  booktitle = {ICML 2026 Workshop on Combining Theory and Benchmarks: Towards
               A Virtuous Cycle to Understand and Guarantee Foundation Model
               Performance},
  year      = {2026},
  url       = {https://openreview.net/forum?id=wHPdBWy4xd}
}

@inproceedings{270750,
  author    = {Sean Quinlan and Sean Dorward},
  title     = {Venti: A New Approach to Archival Data Storage},
  booktitle = {Conference on File and Storage Technologies (FAST 02)},
  year      = {2002},
  address   = {Monterey, CA},
  url       = {
               https://www.usenix.org/conference/fast-02/venti-new-approach-archival-data-storage
               },
  publisher = {USENIX Association},
  month     = jan
}

@inproceedings{1404014.1404025,
  author    = {Rhea, Sean and Cox, Russ and Pesterev, Alex},
  title     = {Fast, inexpensive content-addressed storage in foundation},
  year      = {2008},
  publisher = {USENIX Association},
  address   = {USA},
  booktitle = {USENIX 2008 Annual Technical Conference},
  pages     = {143–156},
  numpages  = {14},
  location  = {Boston, Massachusetts},
  series    = {ATC'08}
}

@article{10.1145/502059.502054,
  author     = {Dabek, Frank and Kaashoek, M. Frans and Karger, David and Morris,
                Robert and Stoica, Ion},
  title      = {Wide-area cooperative storage with CFS},
  year       = {2001},
  issue_date = {Dec. 2001},
  publisher  = {Association for Computing Machinery},
  address    = {New York, NY, USA},
  volume     = {35},
  number     = {5},
  issn       = {0163-5980},
  url        = {https://doi.org/10.1145/502059.502054},
  doi        = {10.1145/502059.502054},
  journal    = {SIGOPS Oper. Syst. Rev.},
  month      = oct,
  pages      = {202–215},
  numpages   = {14}
}

@inproceedings{271167,
  author    = {Kevin Fu and M. Frans Kaashoek and David Mazi{\`e}res},
  title     = {Fast and Secure Distributed {Read-Only} File System},
  booktitle = {Fourth Symposium on Operating Systems Design and Implementation
               (OSDI 2000)},
  year      = {2000},
  address   = {San Diego, CA },
  url       = {
               https://www.usenix.org/conference/osdi-2000/fast-and-secure-distributed-read-only-file-system
               },
  publisher = {USENIX Association},
  month     = oct
}

@inproceedings{1267359.1267365,
  author    = {Nath, Partho and Kozuch, Michael A. and O'Hallaron, David R. and
               Harkes, Jan and Satyanarayanan, M. and Tolia, Niraj and Toups, Matt
               },
  title     = {Design tradeoffs in applying content addressable storage to
               enterprise-scale systems based on virtual machines},
  year      = {2006},
  publisher = {USENIX Association},
  address   = {USA},
  booktitle = {Proceedings of the Annual Conference on USENIX '06 Annual
               Technical Conference},
  pages     = {6},
  numpages  = {1},
  location  = {Boston, MA},
  series    = {ATEC '06}
}

@article{10.1145/502059.502052,
  author     = {Muthitacharoen, Athicha and Chen, Benjie and Mazi{\`e}res, David},
  title      = {A low-bandwidth network file system},
  year       = {2001},
  issue_date = {Dec. 2001},
  publisher  = {Association for Computing Machinery},
  address    = {New York, NY, USA},
  volume     = {35},
  number     = {5},
  issn       = {0163-5980},
  url        = {https://doi.org/10.1145/502059.502052},
  doi        = {10.1145/502059.502052},
  journal    = {SIGOPS Oper. Syst. Rev.},
  month      = oct,
  pages      = {174–187},
  numpages   = {14}
}

@inproceedings{10.1145/2517349.2522721,
  author    = {Mashtizadeh, Ali Jos{\'e} and Bittau, Andrea and Huang, Yifeng
               Frank and Mazi{\`e}res, David},
  title     = {Replication, history, and grafting in the Ori file system},
  year      = {2013},
  isbn      = {9781450323888},
  publisher = {Association for Computing Machinery},
  address   = {New York, NY, USA},
  url       = {https://doi.org/10.1145/2517349.2522721},
  doi       = {10.1145/2517349.2522721},
  booktitle = {Proceedings of the Twenty-Fourth ACM Symposium on Operating
               Systems Principles},
  pages     = {151–166},
  numpages  = {16},
  location  = {Farminton, Pennsylvania},
  series    = {SOSP '13}
}

@inproceedings{10.1145/3379597.3387510,
  author    = {Pietri, Antoine and Spinellis, Diomidis and Zacchiroli, Stefano},
  title     = {The Software Heritage Graph Dataset: Large-scale Analysis of Public
               Software Development History},
  year      = {2020},
  isbn      = {9781450375177},
  publisher = {Association for Computing Machinery},
  address   = {New York, NY, USA},
  url       = {https://doi.org/10.1145/3379597.3387510},
  doi       = {10.1145/3379597.3387510},
  booktitle = {Proceedings of the 17th International Conference on Mining
               Software Repositories},
  pages     = {1–5},
  numpages  = {5},
  location  = {Seoul, Republic of Korea},
  series    = {MSR '20}
}

@inproceedings{10.1145/3544216.3544232,
  author    = {Trautwein, Dennis and Raman, Aravindh and Tyson, Gareth and Castro
               , Ignacio and Scott, Will and Schubotz, Moritz and Gipp, Bela and
               Psaras, Yiannis},
  title     = {Design and evaluation of IPFS: a storage layer for the
               decentralized web},
  year      = {2022},
  isbn      = {9781450394208},
  publisher = {Association for Computing Machinery},
  address   = {New York, NY, USA},
  url       = {https://doi.org/10.1145/3544216.3544232},
  doi       = {10.1145/3544216.3544232},
  booktitle = {Proceedings of the ACM SIGCOMM 2022 Conference},
  pages     = {739–752},
  numpages  = {14},
  location  = {Amsterdam, Netherlands},
  series    = {SIGCOMM '22}
}

@online{ostree,
  author       = {{OSTree}},
  year         = {2026},
  title        = {{Anatomy of an OSTree repository}},
  url          = {https://ostreedev.github.io/ostree/repo/},
  lastaccessed = {September 21, 2026}
}

@inproceedings{269552,
  author    = {Eelco Dolstra and Merijn de Jonge},
  title     = {Nix: A Safe and {Policy-Free} System for Software Deployment },
  booktitle = {18th Large Installation System Administration Conference (LISA
               04)},
  year      = {2004},
  address   = { Atlanta, GA},
  url       = {
               https://www.usenix.org/conference/lisa-04/nix-safe-and-policy-free-system-software-deployment
               },
  publisher = {USENIX Association},
  month     = nov
}

@inproceedings{10.1145/1411204.1411255,
  author    = {Dolstra, Eelco and L{\"o}h, Andres},
  title     = {NixOS: a purely functional Linux distribution},
  year      = {2008},
  isbn      = {9781595939197},
  publisher = {Association for Computing Machinery},
  address   = {New York, NY, USA},
  url       = {https://doi.org/10.1145/1411204.1411255},
  doi       = {10.1145/1411204.1411255},
  booktitle = {Proceedings of the 13th ACM SIGPLAN International Conference on
               Functional Programming},
  pages     = {367–378},
  numpages  = {12},
  location  = {Victoria, BC, Canada},
  series    = {ICFP '08}
}

@misc{courtès2013functionalpackagemanagementguix,
  title         = {Functional Package Management with Guix},
  author        = {Ludovic Courtès},
  year          = {2013},
  eprint        = {1305.4584},
  archiveprefix = {arXiv},
  primaryclass  = {cs.PL},
  url           = {https://arxiv.org/abs/1305.4584}
}

@inproceedings{ORBi-3df206e9-d0b2-4a53-99ed-66d67c2946a3,
  author       = {Cardoen, Guillaume and Mens, Tom and Decan, Alexandre},
  eprint       = {https://orbi.umons.ac.be/handle/20.500.12907/56423},
  eprinttype   = {hdl},
  title        = {GHAgentFiles: A dataset of coding agent file histories in GitHub
                  repositories},
  language     = {English},
  year         = {16 September 2026},
  publisher    = {Institute of Electrical and Electronics Engineers (IEEE)},
  booktitle    = {Proceedings of the 42nd International Conference on Software
                  Maintenance and Evolution},
  location     = {Benevento, Italy},
  organization = {F.R.S.-FNRS - Fonds de la Recherche Scientifique},
  note         = {This dataset paper has received the Distinguished Paper Award for
                  the Tool Demonstration and Data Showcase Track at the 2026 IEEE
                  International Conference on Software Maintenance and Evolution
                  (ICSME)},
  url          = {https://doi.org/10.1109/ICSME67971.2026.00080}
}

@misc{yao2022react,
      title={ReAct: Synergizing Reasoning and Acting in Language Models}, 
      author={Shunyu Yao and Jeffrey Zhao and Dian Yu and Nan Du and Izhak Shafran and Karthik Narasimhan and Yuan Cao},
      year={2023},
      eprint={2210.03629},
      archivePrefix={arXiv},
      primaryClass={cs.CL},
      url={https://arxiv.org/abs/2210.03629}, 
}

@inproceedings{646752.704751,
  author    = {Merkle, Ralph C.},
  title     = {A Digital Signature Based on a Conventional Encryption Function},
  year      = {1987},
  isbn      = {3540187960},
  publisher = {Springer-Verlag},
  address   = {Berlin, Heidelberg},
  booktitle = {A Conference on the Theory and Applications of Cryptographic
               Techniques on Advances in Cryptology},
  pages     = {369–378},
  numpages  = {10},
  series    = {CRYPTO '87}
}

@inproceedings{jimenez2024swebench,
  author    = {Jimenez, Carlos E. and Yang, John and Wettig, Alexander and
               Yao, Shunyu and Pei, Kexin and Press, Ofir and
               Narasimhan, Karthik R.},
  title     = {{SWE}-bench: Can Language Models Resolve Real-World
               {GitHub} Issues?},
  booktitle = {Proceedings of the Twelfth International Conference on
               Learning Representations},
  year      = {2024},
  month     = may,
  address   = {Vienna, Austria},
  publisher = {OpenReview.net},
  url       = {https://openreview.net/forum?id=VTF8yNQM66}
}

@online{aider-edit-formats,
  author       = {{Aider}},
  year         = {2026},
  title        = {{Edit Formats}},
  url          = {https://aider.chat/docs/more/edit-formats.html},
  lastaccessed = {September 25, 2026}
}

\appendix

\clearpage
\setcounter{page}{1}

\end{document}